%% file: main.tex
\documentclass[journal,twoside,web,dvipsnames]{ieeecolor}
\usepackage{generic}
\usepackage{cite}
\usepackage{amsmath,amssymb,amsfonts,mathrsfs}
\usepackage{algorithm,algorithmic}
\usepackage{graphicx}
\usepackage{textcomp}
\usepackage{gensymb}
\usepackage{theorem}
\usepackage{dsfont}
\usepackage{xcolor,color}
\usepackage{hyperref}
\usepackage[draft]{todonotes}
\usepackage{tikz}
\usepackage{caption}
\usepackage{subcaption}
\usepackage{pgfplots,pgfplotstable}
\usetikzlibrary{pgfplots.groupplots}
\pgfplotsset{compat=1.4}
\usepackage{pgf}
\allowdisplaybreaks[4]
\usepackage{multirow}
\usepackage{booktabs}
\usepgfplotslibrary{dateplot,fillbetween}
\usepgfplotslibrary{colorbrewer}
\def\BibTeX{{\rm B\kern-.05em{\sc i\kern-.025em b}\kern-.08em
T\kern-.1667em\lower.7ex\hbox{E}\kern-.125emX}}

\definecolor{MyRed}{HTML}{e6550d}
\definecolor{EPFLRed}{HTML}{b51f1f}
\definecolor{MyBlue}{rgb}{0.20, 0.6, 0.78}
\definecolor{MyGreen}{rgb}{0.4,0.8,0.4}

\newcommand{\change}[1]{\textcolor{black}{#1}}
\newcommand{\newchange}[1]{\textcolor{black}{#1}}

\newtheorem{lemma}{Lemma}
\newtheorem{theorem}{Theorem}
\newtheorem{definition}{Definition}
\newtheorem{remark}{Remark}

\begin{document}
\title{\LARGE \bf A Proximal-Point Lagrangian Based Parallelizable Nonconvex Solver for Bilinear Model Predictive Control
}
\author{Yingzhao Lian,~\IEEEmembership{Graduate Student Member,~IEEE}, Yuning Jiang,~\IEEEmembership{Member,~IEEE}, \\
Daniel F. Opila,~\IEEEmembership{Senior Member,~IEEE}, and Colin N. Jones,~\IEEEmembership{Senior Member,~IEEE}
\thanks{This work has received support from the Swiss National Science Foundation under the RISK project (Risk Aware Data-Driven Demand Response, grant number 200021 175627) and the NCCR Automation project (grant agreement 51NF40\_180545). (Corresponding author: Yuning Jiang)}
\thanks{Yingzhao Lian, Yuning Jiang, and Colin N. Jones are with the Automatic Control Lab, EPFL, Switzerland.
(e-mail: {\tt yingzhao.lian, colin.jones@epfl.ch, yuning.jiang@ieee.org})}
\thanks{Daniel F. Oplia is with the United States Naval Academy, Electrical and Computer Engineering Department. (email: {\tt opila@usna.edu})}}

\maketitle

\begin{abstract}
Nonlinear model predictive control has been widely adopted to manipulate bilinear systems with dynamics that include products of the inputs and the states. These systems are ubiquitous in chemical processes, mechanical systems, and quantum physics, to name a few. Running a bilinear MPC controller in real time requires solving a non-convex optimization problem within a limited sampling time. This paper proposes a novel parallel proximal-point Lagrangian based bilinear MPC solver via an interlacing horizon-splitting scheme. The resulting algorithm converts the non-convex MPC control problem into a set of parallelizable small-scale multi-parametric quadratic programs (mpQPs) and an equality-constrained linear-quadratic regulator problem. As a result, the solutions of mpQPs can be pre-computed offline to enable efficient online computation. The proposed algorithm is validated on a simulation of an HVAC system control. It is deployed on a TI LaunchPad  XL F28379D microcontroller to execute speed control on a field-controlled DC motor, where the MPC updates at 10 ms and solves the problem in 1.764 ms on average and at most 2.088 ms.
\end{abstract}

\section{Introduction}\label{sect:intro}
Bilinear systems were originally introduced in~\cite{mohler1968multivariable,rink1968completely} to model systems where the dynamics involve products of the inputs and the states. These dynamics may result from linearizing a nonlinear input affine system and are most commonly used to model convection and spinning in chemical processes and mechanical systems~\cite{ruberti1974variable,elliott2001bilinear}. Additionally, using the concept of Carleman linearization~\cite{steeb1989note}, it has been shown that bilinear systems can model general nonlinear systems~\cite{kaiser2020data}. Meanwhile, with the help of various sophisticated tools such as Lie algebra~\cite[Chapter 2]{elliott2009bilinear} and Volterra series~\cite{rugh1981nonlinear}, bilinear control theory has been explored in-depth and has found various successful applications~\cite{baillieul1978geometric,rajguru2004three,d2001optimal,escobar1999experimental,sira1987sliding}.

Nonlinear model predictive control (NMPC) is one of the most successful approaches to control bilinear systems~\cite{peitz2018controlling,haddad2012stability,kane2014model,hetel2015binary}. The main idea of NMPC is to achieve the desired performance by optimizing the input in a receding horizon scheme while enforcing state and input constraints~\cite{rawlings2017model}. This requires the solution of a nonlinear optimal control problem (OCP) online within a limited update time. Therefore, an efficient solver \newchange{is critical to running} NMPC in real time\footnote{\change{In this work, real-time means that the MPC solver should return the solution fast enough to enable a desirable operation of the targeted system. Based on our experience, for a mechatronic/mechanical system, the MPC solver should be at least five times faster than the sampling frequency.}}. 

Among various real-time NMPC methods, designing and executing an online solver that can run in parallel via distributed algorithms has been a trend over the past decade~\cite{boyd2011distributed}. Compared to centralized solution approaches, parallelizable methods split the problem into multiple smaller problems such that the computational resources can be utilized more efficiently by exploiting the structure of the OCP being solved. A classical approach used in distributed optimization is based on dual decomposition, where, for example, a gradient-based method~\cite{rantzer2009dynamic,necoara2009interior} or a semi-smooth Newton method~\cite{frasch2015parallel} have been used to solve the concave dual problem. Another famous approach is the Alternating Direction Method of Multipliers, which parallelizes the computation by introducing auxiliary variables~\cite{richter2011towards,boyd2011distributed}. These two methods lack convergence guarantees for nonconvex problems and hence are only formally applicable to linear systems. In~\cite{houska2016augmented}, an augmented Lagrangian based distributed optimization algorithm is proposed, which has been applied to parallelize the computation of MPC problems in~\cite{jiang2020parallel,jiang2019time}. However, despite being parallelized, these algorithms require a solution to multiple non-convex optimization problems in each iteration, which are still numerically intense. 

Decomposing an NMPC problem into a set of small-scale problems mainly leverages the linear equality constraints that appear in NMPC problems, which can reflect the topology of a network system or that naturally emerge in the temporal direction via the introduction of auxiliary variables. The latter approach is the horizon splitting method~\cite{laine2019parallelizing,deng2019parallel}, or sometimes termed Schwarz decomposition~\cite{shin2019parallel}. It splits the predictive trajectories into short sequential sequences, where linear couplings naturally enforce the equality between the initial and terminal states of two adjacent short sequences, hence the name. 
Within the scope of horizon splitting, tools beyond distributed algorithms have been leveraged to improve efficiency further. The banded structure of the KKT system is the most investigated object in this setup.  A binary-tree-structured algorithm summarizes In~\cite{nielsen2015parallel}, a general parallel solver, and in~\cite{deng2019parallel}, an approximation scheme is introduced to develop a parallel Ricatti solver. However, these algorithms still handle the nonconvex problem directly and, as such, are still numerically challenging. 

Another category of methods widely used in real-time NMPC leverages the super-linear local convergence property of Newton-type methods to accelerate online convergence, given a good initialization of the decision variables. This category of methods roughly defines the ``warm-start" strategy, whose initialization usually derives from the solution information gathered from the preceding time step. A basic approach directly shifts the solution from the last iteration~\cite{wang2009fast}, and then a Newton iteration ensures efficient local convergence. Under the umbrella of sequential quadratic programming (SQP), the sensitivity information of the local solution is further used to initialize the KKT system, where an initial guess of active constraints is the most challenging object. In~\cite{ferreau2008online}, the piece-wise affine property of linear \change{model predictive control (MPC)} is used to estimate the change of the active constraint. This idea is generalized in~\cite{diehl2005real} under the name of real-time iteration \change{(RTI)}, where a sensitivity analysis of the local solution is used to give a piece-wise affine update of the control law. 

Instead of solving the NMPC directly online, explicit MPC shifts the online computational burden offline. It treats the MPC control law as a nonlinear mapping from the initial state to control input, and this control law is precomputed offline to enable efficient online calls. In a linear MPC setup, the optimal control law is locally affine~\cite{zafiriou1990robust,bemporad2002explicit}, and this piece-wise affine parametric solution is first used to pre-compute the MPC control law offline in~\cite{bemporad2002explicit}. However, this algebraic property only holds for linear systems, and its application to nonlinear MPC is limited without approximation~\cite{raimondo2011robust}.

This work proposes a new proximal-point Lagrangian based algorithm, which combines the ideas of horizon splitting, explicit MPC, and real-time SQP. In contrast to a standard horizon-splitting approach, a novel interlacing horizon-splitting scheme is introduced. The advantages of the proposed controller are summarized as follows:
\begin{enumerate}
\item \change{The proposed algorithm runs computationally efficient iterations, which only require an evaluation of a multi-parametric QP (mpQP) solution and to solve a sparse linear equation system.}
\item \change{The detection of the active set is shifted to the mpQP solution, whose problem size is independent of the prediction horizon.}
\item A novel interlacing horizon splitting scheme is introduced. The resulting problem has the same number of decision variables as the original NMPC problem without introducing auxiliary variables.
\item \change{The proposed algorithm will not abort even when an infeasible initial state is given. It will output a solution that at least satisfies the input constraint.}
\end{enumerate}

After introducing notation and background knowledge in the rest of this Section~\ref{sect:intro}, the bilinear MPC control problem is presented in Section~\ref{sect:pre}, after which the parallelizable non-convex solver is proposed in Section~\ref{sect:main}. In particular, Section~\ref{sect:sect_split} introduces a novel interlacing horizon splitting scheme, based on which the solver is detailed in Section~\ref{sect:param}. Convergence properties of the proposed solver are studied in Section~\ref{sect:convergence}. \change{After introducing the proposed algorithm, a dual space interpretation of the proposed algorithm is given in Section~\ref{sect:dual}, after which \newchange{a comparison with related results follows} in Section~\ref{sect:compare}. The numerical details of the proposed algorithm are investigated in Section~\ref{sec::impl}. The efficacy of the proposed algorithm is studied in Section~\ref{sect:num}, where the efficient real-time MPC solver \texttt{acados}  and the nonconvex parallel primal-dual solver augmented Lagrangian based alternating direction inexact Newton (ALADIN) method are used as a benchmark. Meanwhile, the proposed algorithm is deployed on a Texas Instruments C2000 Delfino LaunchPad XL F28379 microcontroller to control a field-controlled DC motor.}

\subsection{Notation} 
We use the symbols $\mathbb S_{+}^{n}$ and $\mathbb S_{++}^{n}$ to denote the set of symmetric, positive semi-definite, and symmetric, positive definite matrices in $\mathbb R^{n \times n}$. For a given matrix \change{$\Sigma \in \mathbb S_{++}^{n}$} the notation
$
\left\| x \right\|_{\Sigma} = \sqrt{ x^\top \Sigma x }
$
is used, \change{and $\lVert x\rVert$ denotes the Euclidean norm.} Moreover, a function $c: \mathbb R^{n} \to \mathbb R \cup \{ \infty \}$ is called strongly convex with matrix parameter $\Sigma \in \mathbb S_{+}^n$, if the inequality
\[
c(t x + (1-t)y) \leq t c(x) + (1-t)c(y) - \frac{1}{2} t (1-t) \left\| x-y \right\|_{\Sigma}^2
\]
is satisfied for all $x,y \in \mathbb R^{n}$ and all $t \in [0,1]$. Notice that all convex functions in this paper are assumed to be closed and proper~\cite{Boyd2004}. For a vector $x\in\mathbb{R}^{n}$, we denote by $[x]_i$ its $i$-th element. Set $\mathbb{Z}_i^j$ denotes the range of integers from $i$ to $j$ with $i\leq j$. \change{The Kronecker product of two matrices $A \in \mathbb{R}^{k \times l}$ and $B \in \mathbb{R}^{m \times n}$ is:
\begin{align*}
    \mathbb{R}^{km\times ln}\ni A\otimes B:= \begin{bmatrix}
    a_{1,1}B & a_{1,2}B&\cdots&a_{1,l}B\\
    \vdots & \vdots &\ddots & \vdots\\
    a_{k,1}B & a_{k,2}B &\cdots & a_{k,l}B
    \end{bmatrix}\;.
\end{align*}
$\mathrm{vec}(A)$ denotes the vector that is obtained by stacking all columns of $A$ into one long vector. The reverse operation is denoted by $\mathrm{mat}(\cdot)$, such that $\mathrm{mat}(\mathrm{vec}(A)) = A$.} The identity matrix in $\mathbb{R}^{n\times n}$ is denoted by~$\mathbf I_n$ and the zero matrix in $\mathbb R^{m \times n}$ is denoted by~$\mathbf 0_{m\times n}$. Notation $\mathrm{diag}(H_1,\dots,H_n)$ constructs a block-diagonal matrix whose $i$-th diagonal block is $H_i$. \change{For a given function $f : \mathbb{R}^n \to \mathbb{R}$, we use the Landau notation} 
\[
\change{
f(x) = \mathbf{O}(\left\| x \right\|)\;,\quad \text{if}\;\; \exists\, c \in \mathbb{R}\;,\;\lim_{x \to 0} \frac{f(x)}{\left\| x \right\|} = c\;.}
\]

\subsection{Preliminaries} 
\label{sec::Pre}
We first recap some existing results from the field of multi-parametric quadratic programming (mpQP) used later in this paper. A generic convex mpQP can be written in the form of
\begin{subequations}
\label{eq:pqp}
\begin{align}
\min_x  &\;\; \frac{1}{2} x^{\top} Q x + \theta^{\top} S x  \label{eq:pqp:cost}\\
\text{s.t.} \;&\;\; A x \leq b + C \theta\enspace ,\label{eq:pqp:cons}
\end{align}
\end{subequations}
with decision variables $x \in \mathbb{R}^{n_x}$ and parameters $\theta \in \mathbb{R}^{n_p}$. Here, matrices $Q\in\mathbb{S}_+^{n_x}$, $S\in\mathbb{R}^{n_p\times n_x}$, $A\in\mathbb{R}^{m\times n_x}$, $C\in\mathbb{R}^{m\times n_p}$ and vector $b\in\mathbb{R}^m$ are given data. Moreover, we denote by $\Omega$ the set of all parameters $\theta$ for which~\eqref{eq:pqp} is feasible. For a mpQP~\eqref{eq:pqp} with a strongly convex value function, it has been shown (see, e.g.,~\cite{BemEtal:aut:02}) that $\Omega$ is a polyhedron 
while the solution map $x^\star(\theta):\mathbb R^{n_p}\to \mathbb R^{n_x}$ is a continuous piecewise affine (PWA) function of the parameters. \change{Each affine piece is called a critical region~\cite[Chapter 7.1.2]{borrelli2017predictive}.} Meanwhile, the Lipschitz-continuity holds at $x^\star(\cdot)$, i.e., there exists a positive constant $\eta>0$ such that for any $\theta_1,\theta_2\in\Omega$, we have 
\begin{equation}\label{eq::xLipschitz}
\left\|x^\star(\theta_1) - x^\star(\theta_2)\right\| \leq \eta \left\|\theta_1-\theta_2\right\|.
\end{equation}

We now recall some definitions from the field of nonlinear programming (NLP). Let us consider NLPs in a generic form 
\begin{equation}\label{eq::NLP}
\min_{x} \;\;f(x)\quad\text{subject to}\left\{
\begin{aligned}
g(x) &= 0 \quad \mid \lambda, \\
h(x) &\leq 0 \quad \mid \kappa.
\end{aligned}
\right.
\end{equation}
Throughout the rest of this paper, we write Lagrangian multipliers right after the constraints such that $\lambda\in\mathbb R^{n_g}$ and $\mathbb R^{n_h}\ni\kappa\geq 0 $ denote, respectively, the Lagrangian multipliers of the equality constraints and inequality constraints.
Functions $f:\mathbb R^{n_x}\to \mathbb R$, $g:\mathbb R^{n_x}\to\mathbb R^{n_g}$ and $h:\mathbb R^{n_x}\to\mathbb R^{n_h}$ are assumed twice continuously differentiable. 
A primal-dual solution $(x^*,\lambda^*,\kappa^*)$ is called a Karush–Kuhn–Tucker (KKT) point of~\eqref{eq::NLP} if the following conditions are satisfied~\cite[Chapter 12.3]{Nocedal2006}
\begin{subequations}\label{eqn:kkt_NLP}
\begin{align}
\nabla f(x^*) + \nabla g(x^*)\lambda^* + \nabla h(x^*)\kappa^*&=0,\\
g(x^*) = 0,\;h(x^*)&\leq 0,\\
\forall\,i\in 1,...,n_h,\;\;[\kappa^*]_i \cdot [h(x^*)]_i=0, [\kappa^*]_i &\geq 0,
\end{align}    
\end{subequations}
where $[\kappa^*]_i$ and $[h(x^*)]_i$ define the $i$-th element of $\kappa^*$ and $h(x^*)$, respectively. \change{For a given feasible $x$, we denote by $\mathcal{A}(x)$ the active set at $x$, i.e.,  the index set that includes the equality constraints and the inequality constraints that holds equality at $x$. When the set of active constraint gradients (i.e., $\begin{bmatrix}
\nabla g(x),\nabla h_{i\in\mathcal{A}(x)}(x)\end{bmatrix}$) is linearly independent at point $x$, the linear constraint qualification (LICQ) holds~\cite[Chapter 12.2]{Nocedal2006}. Furthermore, we say the second-order sufficient condition (SOSC) holds at point $x$ if its hessian $\nabla^2 h(x)$ is positive definite semidefinite on the null space spanned by active constraint gradients~\cite{robinson2009second}. Finally, we say the strict complementary condition (SCC) holds if a dual variable equals zero only when the corresponding constraint is \newchange{inactive~\cite[Definition 12.5]{Nocedal2006}}.} 
Then, we state the definition of regular KKT point for NLP~\eqref{eq::NLP}.
\begin{definition}\label{def::regularKKT}
\cite{robinson2009second} A given KKT point $(x^*,\lambda^*,\kappa^*)$ is called a regular KKT point if the LICQ, SOSC, and the SCC hold. 
\end{definition}

For a given KKT point $(x^*,\lambda^*,\kappa^*)$, if it is regular, then there exists an open neighborhood $\mathcal B(x^*)$ around $x^*$ such that the active set is fixed for any $x\in\mathcal{B}(x^*)$, (i.e., $\mathcal A(x)=\mathcal{A}(x^*)$)~\cite{robinson2009second}).  Regularity at KKT points guarantees the local convergence property when a Newton-type method is applied to solve~\eqref{eq::NLP}~\cite{Nocedal2006}. 

\change{
When the inequality constraint $h(x)\leq 0$ defines a convex set, the first-order optimality condition~\eqref{eqn:kkt_NLP} can be further simplified for the sake of compactness:
\[
0\in \nabla f(x^*) + \nabla g(x^*)\lambda^* +\mathcal N_{\mathcal X}(x^*),
\]
with the convex set $\mathcal X:=\{x\in\mathbb R^{n_x} | h(x)\leq 0\} $ and $\mathcal N_{\mathcal X}(x^*):=\left\{y\middle| (y-x^*)^\top (x-x^*)\leq 0,\;\forall\;x\in\mathcal{X}\right\}$ denotes the normal cone of convex set $\mathcal X$ at $x^*$.}

In contrast to the standard Hestenes-Powell augmented Lagrangian method~\cite{hestenes1969multiplier,powell1969method}, a variant of an augmented Lagrangian method, termed proximal-point Lagrangian~\cite{rockafellar1976augmented}, is used in this work. Given an equality constraint optimization problem
\begin{align*}
    \min_{x} \;\;f(x)\quad\text{s.t.}\quad g(x)=0\;,
\end{align*}
its linearized proximal-point Lagrangian around $\overline{x}$ \change{is} defined by 
\begin{align}\label{eq::exAL}
\mathcal{L}^\rho(x,\lambda,\overline{x}):= f(x)+(\nabla g(\overline{x})\lambda)^\top x+\frac{\rho}{2}\lVert x-\overline{x}\rVert^2\;.
\end{align}
Note that when $\rho=0$, it recovers the linearized Lagrangian. \change{For the sake of completeness, the Hestenes-Powell augmented Lagrangian is defined by $f(x)+\lambda^\top g(x)+\frac{\rho}{2}\lVert g(x)\rVert^2$.} 

\section{Problem Formulation}
\label{sect:pre}
This paper considers discrete-time, time-invariant bilinear dynamic systems:
\begin{align}\label{eq::dyn}
x_{k+1} =& Ax_k+Bu_k+\sum_{i=1}^{n_u}C_{i} x_k[u_{k}]_i+ B_w w_k
\end{align}
with state $x_k\in\mathbb R^{n_x}$, control inputs $u_k\in\mathbb{R}^{n_u}$ and disturbance $w_k$ at time instant $k$. For the sake of simplicity, we group the bilinear coefficient matrices $C=[C_1^\top,...,C_{n_u}^\top]^\top\in\mathbb{R}^{n_x\cdot n_u\times n_x}$ and assume that the states and control inputs are subject to the polyhedral constraints
\[
\begin{aligned}
x_k\in\mathcal{X}&:=\left\{x\in\mathbb{R}^{n_x}\left|P_x x\leq p_x\right\}\right.,\\
\text{and}\;\;u_k\in\mathcal{U}&:=\left\{u\in\mathbb{R}^{n_u}\left|P_u u\leq p_u\right\}\right..
\end{aligned}
\]

\change{The dynamics~\eqref{eq::dyn} also includes the case without process noise. For the case with process noise, to make the problem tractable, we consider solving a certainty equivalent problem, where the prediction of the nominal process noise $w_k$ is available. This assumption holds in many energy-related applications: solar radiation in photovoltaic power systems, the outdoor climate in building control, and power generation in airborne wind energy systems, to name a few.} An MPC controller can be designed by recursively solving the following optimal control problem in a receding horizon fashion,
\begin{subequations}\label{eqn:prob_glb}
\begin{align}
\min_{x_0,x,u}\;\;\sum_{k=0}^{N-1} \ell_k(x_k,u_k)+\ell_N(x_N)
\end{align}
subject to:
\begin{align}
& x_0 = x(t),&\\\notag
&\forall\, k\in\{0,1,...,N-1\},&\\\label{eq::dynProb}
&x_{k+1}=Ax_k + Bu_k + \sum_{i=1}^{n_u}C_{i} x_k[u_{k}]_i+B_ww_k\;\; &\mid \lambda_k,\\
& x_{k+1}\in \mathcal{X},\;u_k\in \mathcal{U}\;\; &\mid \kappa_k,
\end{align}
\end{subequations}
with $x=[x_1^\top,...,x_N^\top]^\top$, $u=[u_0^\top,...,u_{N-1}^\top]^\top$ and prediction horizon $N\in\mathbb{Z}_{>0}$. \change{For the sake of consistency, variables indexed by bracketed time, such as $x(t)$, denote the actual measurements read out from sensors. Meanwhile, variables indexed by a subscript, such as $x_k$, denote the predictive ``virtual" variables used in the MPC problem.} In problem~\eqref{eqn:prob_glb}, the stage cost $\ell_k(\cdot,\cdot)$, $k\in\mathbb Z_0^{N-1}$ and terminal cost $\ell_N(\cdot)$ are quadratic and strongly convex, i.e.,
\[
\begin{aligned}
\ell_k(x,u)=& \frac{1}{2}x^\top Q_k x + q_k^\top x + \frac{1}{2}u^\top R_k u + r_k^\top u,\\
\ell_N(x)=& \frac{1}{2}x^\top Q_N x + q_N^\top x
\end{aligned}
\]
with user-defined parameters $Q,Q_N\in\mathbb{S}_{++}^{n_x},\;R\in\mathbb{S}^{n_u}_{++}$ \change{$q_k\in\mathbb{R}^{n_x},\;r_k\in\mathbb{R}^{n_u}$}. Notice that although its objective is strongly convex, solving \newchange{the} nonconvex Problem~\eqref{eqn:prob_glb} is challenging due to the bilinear dynamics~\eqref{eq::dynProb}.

\section{Algorithm Development}
\label{sect:main}
\change{In this work, we propose an algorithm to solve the bilinear MPC problem~\eqref{eqn:prob_glb} efficiently. Before delving into the algorithmic details, we would first state the logic behind the design of the proposed algorithm. As reviewed in Section~\ref{sect:intro}, real-time MPC mainly applies an SQP solver with warm-start strategies or uses explicit MPC. When a good initialization is not available, detecting active inequality constraints becomes the major performance bottleneck for the SQP algorithm. This requires the design of sophisticated, active set detection strategies or the use of merit functions (see e.g.~\cite[Chapter 18.2]{Nocedal2006}~\cite[Chapter  2.3]{Verschueren2021}~\cite[Chapter 2.3]{gill2005snopt}). In the worst case, a poor estimate of the active set will lead to an infeasible QP subproblem, which ultimately aborts the progress of the SQP algorithm. On the contrary, the information of the active set is implicitly saved as critical regions in the explicit solution of explicit MPC. However, its application is limited to linear systems (Section~\ref{sect:intro}).}

\change{This work aims at bringing the benefits of explicit MPC to the SQP method in the application of bilinear MPC. In particular, instead of using an explicit MPC, the explicit solution will play the role of an active set detector in this work. In the rest of this section, we will first introduce a novel interlacing horizon splitting scheme, after which the parallelizable parametric nonconvex solver is elaborated. The convergence properties of the proposed solver are studied in Section~\ref{sect:convergence}. An interpretation of the proposed scheme in the dual space is given in Section~\ref{sect:dual}, followed by a comparison with related works in Section~\ref{sect:compare}.}

\subsection{Interlacing horizon splitting reformulation}
\label{sect:sect_split}
\begin{figure}[htbp!]
\centering
\input{grp_fig}
\caption{Visualization of the interlacing horizon splitting.}
\label{fig:grp}
\end{figure}
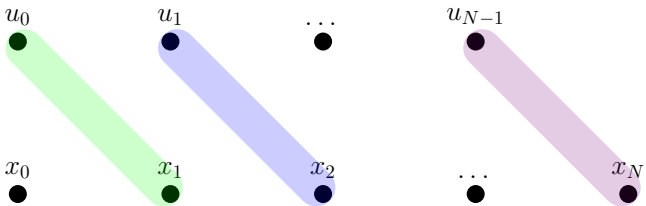
This section presents the interlacing horizon splitting scheme used later to develop a parallelizable parametric solver to deal with~\eqref{eqn:prob_glb}. As depicted in Fig.~\ref{fig:grp}, its main idea is to bind the $k$-th input $u_k$ with state $x_{k+1}$.
To this end, we introduce the shorthand
$\xi_0=x_0$ and $\xi_k =[u_{k-1}^\top,x_{k}^\top]^\top$ for all $k\in\mathbb{Z}_1^{N}$ with associated constraint sets $\Xi_0 = \{\xi\in\mathbb R^{n_x}: \xi_0 =\change{x_0=} x(t) \}$ and 
\begin{align}\label{eqn:cons_combined}
  \begin{split}
\Xi_{k} =& \{\xi \in\mathbb R^{n_u+n_x}: \xi \in \mathcal{U}\times \mathcal{X}\},\;k\in\mathbb Z_1^{N}\\
=& \{\xi \in\mathbb R^{n_u+n_x}: P_\xi\xi \leq p_\xi\} 
\end{split}  
\end{align}
with $P_\xi = \text{diag}(P_u, P_x)$ and $p_\xi=[p_u^\top,p_x^\top]^\top$.
Moreover, we denote the decoupled objective by
\[
\begin{aligned}
F_0(\xi_0)&=\frac{1}{2}\xi_0^\top Q_0\xi_0+q_0^\top \xi_0,\\
F_k(\xi_k) &= \frac{1}{2}\left\|\xi_k\right\|^2_{\text{diag}(R_{k-1},Q_{k})} + [r_{k-1}^\top,q_{k}^\top]^\top \xi_k, \; k\in\mathbb{Z}_1^{N}
\end{aligned}
\]
and summarize the bilinear dynamics~\eqref{eq::dyn} by 
\[
D_k \xi_k + E_k\xi_{k+1} + ( S_{k+1}\xi_{k+1} \otimes \mathbf I_{n_x})^\top G_k\xi_k=d_k
\]
with coefficients $d_k = -B_ww_k,\;k\in\mathbb{Z}_0^{N-1}$,
\[
\begin{aligned}
D_0 &= A,\; D_k = [\mathbf 0_{n_x\times n_u},\;A],\;k\in\mathbb Z_{1}^{N-1}\\
E_k  & =[B,-\mathbf{I}_{n_x}],\;S_k=[\mathbf{I}_{n_u},\;\mathbf{0}_{n_u\times n_x}],\;k\in\mathbb Z_0^{N-1}\\
\text{and}\;\;G_0 &=C,\;G_k = \left[[\mathbf{0}_{n_x\times n_u}, \;C_1]^\top,...,[\mathbf{0}_{n_x\times n_u}, \;C_{n_u}]^\top\right]^\top 
\end{aligned}
\]
for all $k\in\mathbb Z_1^{N-1}$.
Accordingly, Problem~\eqref{eqn:prob_glb} can be rewritten as 
\begin{subequations}\label{eq::prob}
\begin{align}
\min_{\xi}\;&\sum_{k=0}^NF_k(\xi_k)&\\\notag
 \text{s.t.}\;\;& (S_{k+1}\xi_{k+1} \otimes \mathbf I_{n_x})^\top G_k\xi_k&\\\label{eq::probEq}
 &\qquad+ D_k \xi_k + E_k\xi_{k+1}=d_k \;\;&\mid\lambda_k,\;k\in\mathbb Z_0^{N-1}\\
 &\xi_k\in\Xi_k,\;&\mid\kappa_k,\;k\in\mathbb Z_0^{N}\;\;\;.\label{eq::probNeq}
\end{align}
\end{subequations}

\subsection{Proximal-point Lagrangian Based Parallelizable Solver}
\label{sect:param}
\change{Based on the interlacing splitting, the \newchange{linearized proximal-point Lagrangian~\eqref{eq::exAL} is} used to design the algorithm. On the one hand, it gives zero local duality gap even under the nonlinear/non-convex dynamics~\cite[Chapter~11.K]{rockafellar2009variational}. On the other hand, it enables a \textbf{convex} QP, accordingly an mpQP, formulation of parallelizable local problems.} In particular, for a given primal  trajectory $\overline{\xi}$ (i.e inputs and states) and a dual trajectory $\{\lambda_k\}_0^{N-1}$, the linearized proximal-point Lagrangian of~\eqref{eq::prob} w.r.t the equality constraint~\eqref{eq::probEq} is given by 
\begin{align}\label{eq::AL_sum}
   \mathcal{L}^\rho(\xi,\lambda,\overline{\xi} ) =\;& \mathcal{L}_0^\rho(\xi_0,\lambda_0,\overline{\xi}_0,\overline{\xi}_{1})+ \mathcal{L}_N^\rho(\xi_N,\lambda_{N-1},\overline{\xi}_{N-1},\overline{\xi}_{N})\nonumber\\
&+\sum_{k=1}^{N-1} \mathcal{L}_k^\rho(\xi_k,\lambda_{k-1},\lambda_{k},\overline{\xi}_{k-1},\overline{\xi}_{k},\overline{\xi}_{k+1}) 
\end{align}
with
\begin{subequations}
\label{eq::Lrho}
\begin{align}
&\mathcal{L}_0^\rho(\xi_0,\lambda_0,\overline{\xi}_0,\overline{\xi}_{1}):= F_0(\xi_0)\\\notag
&\quad+ \lambda_0^\top \left[D_0+ (S_{1}\overline{\xi}_{1} \otimes \mathbf I_{n_x})^\top G_0\right]\xi_0 + \frac{\rho}{2}\left\|\xi_0-\overline{\xi}_0\right\|^2,\\[0.16cm]
&\mathcal{L}_k^\rho(\xi_k,\lambda_{k-1},\lambda_{k},\overline{\xi}_{k-1},\overline{\xi}_{k},\overline{\xi}_{k+1}):=  \\\notag
&\quad F_k(\xi_k)+ \lambda_{k-1}^\top\left(E_{k-1}+\mathrm{mat}(G_{k-1}\overline{\xi}_{k-1})\cdot S_k\right) \xi_{k}\\\notag
&\quad+ \lambda_k^\top \left[D_k+ (S_{k+1}\overline{\xi}_{k+1} \otimes \mathbf I_{n_x})^\top G_k\right]\xi_k + \frac{\rho}{2}\left\|\xi_k-\overline{\xi}_k\right\|^2,\\[0.16cm]\notag
&\mathcal{L}_N^\rho(\xi_k,\lambda_{N-1},\overline{\xi}_{N-1},\overline{\xi}_{N}):=  F_N(\xi_k)+ \frac{\rho}{2}\left\|\xi_N-\overline{\xi}_N\right\|^2\\
&\quad + \lambda_{N-1}^\top\left(E_{N-1}+\mathrm{mat}(G_{N-1}\overline{\xi}_{N-1})\cdot S_N\right) \xi_{N}
\end{align}
\end{subequations}
with $\rho>0$ and $\lambda:=[\lambda_0^\top,\lambda_1^\top,\dots,\lambda_{N-1}^\top]^\top$. From the proximal-point Lagrangian~\eqref{eq::AL_sum}, the benefits of the interlacing horizon splitting become clear. Firstly, the problem is decomposed into $N+1$ independent subproblems in $\xi$. Secondly, each subproblem is a convex mpQP. Furthermore, the use of the proximal-point Lagrangian allows a simplification of $\mathcal{L}_k^\rho(\cdot)$ to a modified proximal form (Section.~\ref{sect:mpQP})~\cite{di1982new}.

If the primal-dual solution $(\xi^*,\lambda^*)$ of~\eqref{eq::prob} is a regular KKT point, we have solving~\eqref{eq::prob} equivalent to solving 
\begin{equation}\label{eq::dualProb}
\begin{aligned}
\max_\lambda &\left(-\sum_{k=0}^{N-1}\lambda_k^\top d_k + \min_\xi\;\; \mathcal{L}^\rho(\xi,\lambda,\overline{\xi}=\xi^*) \right)\\
&\text{subject to}\;\;\xi\in\Xi = \Xi_0\times ...\times \Xi_N.
\end{aligned}
\end{equation}
As $\mathcal{L}^\rho$ is decoupled in $\xi$, our main idea to develop a parallelizable algorithm solving~\eqref{eq::prob} is to design a primal-dual algorithm to solve the dual problem~\eqref{eq::dualProb} with an iterative update in $\overline{\xi}$. 
\begin{algorithm}[htbp!]
\caption{Proximal-point Lagrangian Based Online Solver for Bilinear MPC}
\textbf{Input:} an initial guess of $(\overline{\xi},\lambda)$, a stop tolerance $\epsilon >0$, a proximal weight $\rho$ and a slack penalty $\mu$ \\
\textbf{Repeat:}
\begin{enumerate}
\item Solve decoupled mpQP problems \textbf{sequentially or in parallel},
\begin{subequations}
\label{eq::deQP}
\begin{align}
&\hspace{-8mm}\min_{\xi_0\in\Xi_0}\;\;\mathcal{L}_0^\rho(\xi_0,\lambda_0,x(t),\overline{\xi}_{1}),\\
&\hspace{-8mm}\min_{\xi_k\in\Xi_k}\;\;\mathcal{L}_k^\rho(\xi_k,\lambda_{k-1},\lambda_{k},\overline{\xi}_{k-1},\overline{\xi}_{k},\overline{\xi}_{k+1}),\;k\in\mathbb Z_1^{N-1},\\
&\hspace{-8mm}\min_{\xi_N\in\Xi_N}\;\mathcal{L}_N^\rho(\xi_N,\lambda_{N-1},\overline{\xi}_{N-1},\overline{\xi}_{N}).
\end{align}
\end{subequations}
In all the following steps, $\xi_k$, $k\in\mathbb Z_0^N$ denote optimal solutions of the above QPs.
\item Evaluate sensitivities
\begin{subequations}
\begin{align}
    &\hspace{-4mm}H \approx \nabla_{\xi\xi} \mathcal{L}^0(\xi,\lambda,\xi),\\
    &\hspace{-4mm}g_k = \nabla F_k(\xi_k) - \nabla \mathcal{L}_k^\rho(\xi,\lambda,\xi)\;,\;k\in\mathbb Z_0^{N},\label{eq::sen_Lagrangian}\\\notag
    &\hspace{-4mm}c_k = D_k \xi_k + E_k\xi_{k+1} \\
    &\qquad\qquad + ( S_{k+1}\xi_{k+1} \otimes \mathbf I_{n_x})^\top G_k\xi_k -d_k,
\end{align}
\end{subequations}
and the active Jacobian $\hat P_\xi^k$ at local solution $\xi_k$. Here, we use simplified notation $\mathcal L^\rho_k(\xi,\lambda,\xi)$, $k\in\mathbb Z_0^N$. for~\eqref{eq::Lrho}. 
\item Terminate if $\max_k\|c_k\|\leq \epsilon$ and $\max_k\rho\|\xi_k-\overline \xi_k\|\leq \epsilon$ hold.

\item Solve equality-constrained QP
\begin{subequations}
\label{eq::coQP}
\begin{align}
    \min_{\Delta \xi, s}\;\;&\frac{1}{2}\Delta \xi^\top H \Delta \xi + \sum_{k=1}^{N} \left\{g_k^\top \Delta \xi_k + \mu\|s_k\|^2\right\}\\
    \text{s.t.}\;\;&\Delta \xi_0 = 0 \\\notag
    &E_k\Delta \xi_{k+1} + ( S_{k+1}\xi_{k+1} \otimes \mathbf I_{n_x})^\top G_k\Delta \xi_k\\\notag
    &\quad +\mathrm{mat}(G_{k}\xi_{k})\cdot S_{k+1}\Delta \xi_{k+1}\\\label{eq::coDyn}
    &\quad +c_k + D_k \Delta \xi_k =0\; \mid \lambda_k^\mathrm{QP},\;k\in\mathbb{Z}_0^{N-1}\\\label{eq::coActive}
    &\hat P_\xi^k \Delta \xi_k = s_k,\;k\in\mathbb Z_1^N.
\end{align}
\end{subequations}
\item Update $\overline{\xi}^+ = \xi +\Delta \xi $ and $\lambda^+=\lambda^\mathrm{QP}$. 
\end{enumerate}
\label{alg::solver}
\end{algorithm}

Algorithm~\ref{alg::solver} outlines the main steps of the proposed algorithm for solving~\eqref{eq::prob}. Step 1) deals with decoupled problem~\eqref{eq::deQP} in parallel, which has explicit solutions as \textbf{convex} mpQPs. Particularly, their solution maps are piece-wise affine functions and can be pre-computed offline (See Section~\ref{sect:mpQP}). Based on the local solutions $\xi$, Step~2) evaluates the sensitivities, including the Hessian approximation of the Lagrangian $\mathcal L^0$, the gradients of the decoupled objective $F_k$ and the bilinear dynamics residual $c_k$. The active Jacobian $\hat P_\xi^k$ are constructed based on the active set at local solutions $\xi_k$. The terminal condition is given in Step~3). It is clear that if these termination conditions hold, we have the iterate $(\xi,\lambda)$ satisfying the \change{first-order optimality condition
\[
\mathbf{O}(\epsilon)
\in \nabla_\xi \mathcal{L}^0(\xi,\lambda,\overline{\xi}) + \mathcal N_{\Xi}(\xi)
\]
with $\Xi=\Xi_0\times \Xi_1 \times \cdots \times \Xi_{N}$} and the primal feasibility condition 
\[
\left\|D_k \xi_k + E_k \xi_{k+1} + ( S_{k+1}\xi_{k+1} \otimes \mathbf I_{n_x})^\top G_k\xi_k\right\|=\mathbf
O(\epsilon)
\]
for all $k\in\mathbb Z_0^{N-1}$ up to a small error of order $\mathbf{O}(\epsilon)$. Step~4) deals with a structured equality-constrained QP~\eqref{eq::coQP}. To overcome the potential infeasibility caused by the linearization of nonlinear dynamic~\eqref{eq::dyn} in constraint~\eqref{eq::coDyn}, we introduce a decoupled slackness $s_k$ for each active constraint~\eqref{eq::coActive}. This makes QP~\eqref{eq::coQP} always feasible regardless of the feasibility of the original problem~\eqref{eq::prob}. \change{Note that step 4) is similar to an SQP step, while the mpQP solutions directly generate its active sets.} On top of this, the mpQPs in step 1) are also always feasible. Therefore, if one applies Algorithm~\ref{alg::solver} as an online solver for MPC, the resulting MPC controller is always feasible, i.e., the iteration of Algorithm~\ref{alg::solver} will never fail before termination, and it is independent of the initial condition $x(t)$. This property is desirable in real-world applications because handling infeasibility requires careful design/tuning of a relaxed problem. \change{Furthermore, even for a feasible problem, the standard SQP algorithm may abort due to an incorrect estimate of the active sets.  More specifically, if the estimated active set leads to an infeasible QP, the iterations of the SQP algorithm will stop. In summary, the interlacing horizon splitting scheme enables the mpQPs formulation. The proposed algorithm thereby iteratively calls the mpQP solutions, and the inputs to the mpQPs are iteratively updated in the SQP-style step 4). }
\begin{remark}
As discussed above, the proposed solver is always feasible even when the initial state makes the NMPC problem infeasible. This property is also observed in the compositions of the projection operator ~\cite{alwadani2021difference,alwadani2021resolvents}, whose convergence to a point pair that are closest to all the sets is proved. However, in a nonconvex setup, the property of the convergent results is unclear and remains open for future research.
\end{remark}

\begin{remark}
\newchange{In Section~\ref{sect:pre}, we discussed that the proposed algorithm only considers process noise in a certainty equivalent form, assuming that the nominal process noise is available. However, the proposed algorithm is not able to provide an efficient solution to the robust NMPC problem, which is a challenging nonconvex problem that requires further investigation in the future. However, it is worth noting that the proposed algorithm still has practical benefits even in the current setup. For instance, in building control, even if weather forecasts are available, the actual weather may not align with the nominal forecast, causing the states of the building to evolve into an initial state that renders the NMPC infeasible. The property that the proposed algorithm remains feasible at all times ensures that the building's operations continue uninterrupted.}
\end{remark}

\subsection{Local Convergence Property}\label{sect:convergence}
\change{This section shows that the proposed Algorithm~\ref{alg::solver} asymptotically converges to the local optimal solution of~\eqref{eqn:prob_glb} at a quadratic rate. The logic behind the constructive proof follows two facts: local mpQPs~\eqref{eq::deQP} have a Lipschitz-continuous solution map, and the coupled QP~\eqref{eq::coQP} is equivalent to a Newton-type method. To this end, we introduce the following lemma first to establish the quadratic contraction of the solution of~\eqref{eq::coQP}.}
\begin{lemma}
\label{lem::contraction}
Let the KKT point $(\xi^*,\lambda^*)$ of Problem~\eqref{eq::prob} be regular such that the solution $\xi^*$ is a local minimizer.  Moreover, \textcolor{black}{let Algorithm~\ref{alg::solver} be initialized locally~\footnote{\textcolor{black}{The term "local" in the statement throughout this paper means that the initial guess of primal-dual iterates is located within a small neighborhood around the local solution $(\xi^*,\lambda^*)$. Hence, the condition~\eqref{eq::HessianCond} is required to be satisfied locally. It is worth mentioning that the assumption of locality is standard and widely used in the local convergence analysis of Newton's type methods~\cite{Nocedal2006}.}}, whose Hessian evaluation $H$ and parameter $\mu$ in~\eqref{eq::coQP} satisfy
\begin{equation}
\label{eq::HessianCond}
H = \nabla_{\xi\xi} \mathcal{L}^0(\xi,\lambda,\xi) + \mathbf{O}(\|\xi-\overline{\xi}\|)\;\;\text{and}\;\;\frac{1}{\mu}\leq \mathbf{O}(\|\xi-\overline{\xi}\|),
\end{equation} 
respectively for every iterate $(\xi,\overline \xi)$.}
\change{Then, there exists $\alpha >0$,} the solution to~\eqref{eq::coQP} locally satisfies,
\begin{equation}
\label{eq::contraction}
\left\lVert \overline{\xi}^+ - \xi^* \right\| \leq \alpha \left\rVert\xi-\xi^*\right \|^2,\; \left\| \lambda ^+ - \lambda^* \right\| \leq \alpha \|\xi-\xi^*\|^2.
\end{equation}
\end{lemma}
\begin{proof}
\change{
Based on the definition~\ref{def::regularKKT} of regular KKT point, we have that the active sets are not changed locally~\cite{Houska2021}. Then, the standard analysis of Newton's method gives 
\[
\begin{aligned}
\left\|\begin{bmatrix}
\overline{\xi}^+-\xi^*\\ \lambda^+-\lambda^*
\end{bmatrix}\right\|\leq& \left\|H-\nabla_{\xi\xi}\mathcal{L}^0(\xi,\lambda,\xi)\right\|\cdot\mathbf{O}\left(\left\|\xi-\xi^*\right\|\right)\\ 
&+\mathbf{O}\left(\left\|\xi-\xi^*\right\|^2\right)\overset{\eqref{eq::HessianCond}}{\leq} \mathbf{O}\left(\left\|\xi-\xi^*\right\|^2\right) 
\end{aligned}
\]
as discussed in~\cite[Chapter 3.3]{Nocedal2006}, which concludes the proof.} 
\end{proof}

\change{Intuitively speaking, this lemma states that if the iterates given by~\eqref{eq::deQP} are close to the optimal solution to~\eqref{eqn:prob_glb}, then the distance between iterate given by~\eqref{eq::coQP} and the optimal solution contract quadratically. The following theorem intends to show that this quadratic contraction holds even when we consider the iterates given by~\eqref{eq::deQP}.} Based on condition~\eqref{eq::HessianCond}, we have that there exists a constant $\alpha>0$ such that the local quadratic contraction~\eqref{eq::contraction} holds. Then, we define by $\xi^+$ the solution of~\eqref{eq::deQP} based on the updated primal-dual iterate $(\overline{\xi}^+,\lambda^+)$ such that we can summarize the local convergence result as follows. 
\begin{theorem}\label{thm:convergence}
\change{Let all assumptions in Lemma~\ref{lem::contraction} be satisfied. \change{The iterates $\xi$} of Algorithm~\ref{alg::solver} locally converges to the local minimizer $\xi^*$ of Problem~\eqref{eq::prob} with quadratic rate.}
\end{theorem}
\begin{proof}
\change{
As discussed in Section~\ref{sec::Pre}, we have the local solution map $\xi^\star(\overline{\xi},\lambda)$ of convex mpQPs~\eqref{eq::deQP} are Lipschitz continuous such that we have 
\[
\begin{aligned}
\left\|\xi^\star(\overline{\xi}^+,\lambda^+)-\xi^\star(\xi^*,\lambda^*)\right\|= \left\|\xi^+ - \xi^*\right\| \leq \eta\left\|
\begin{bmatrix}
\overline{\xi}^+-\xi^*\\ \lambda^+-\lambda^*
\end{bmatrix}
\right\|
\end{aligned}
\]
with a constant $\eta>0$. Here, we use the fact $\xi^*=\xi^\star(\xi^*,\lambda^*)$, i.e., if we initialize Algorithm~\ref{alg::solver} with the optimal solution $(\xi^*,\lambda^*)$, the solution of convex mpQPs~\eqref{eq::deQP} is equal to the local minimizer $\xi^*$. Moreover, iterate $\xi^+$ is the solution of~\eqref{eq::deQP} if one starts the Algorithm~\ref{alg::solver} with $(\overline{\xi}^+,\lambda^+)$ as the initial guess. Substituting~\eqref{eq::contraction} into the inequality above yields
\[
\left\|\xi^+ - \xi^*\right\|\leq \alpha\cdot\eta \left\|\xi - \xi^*\right\|^2,
\]
which is sufficient to establish the local quadratic convergence of iterates $\xi$ to the local minimizer $\xi^*$~\cite[Chapter 3.3]{Nocedal2006}.}
\end{proof} 

\change{It is worth mentioning that the same order of local convergence speed holds in a wide range of second-order algorithms, such as the SQP algorithm~\cite[Chapter 18.7]{Nocedal2006} and the augmented Lagrangian based alternating direction inexact Newton (ALADIN) method~\cite{houska2016augmented}. The theoretical importance of Theorem 1 shows that such convergence rate is still preserved even when another layer of mpQPs is added (i.e., step 1) in Algorithm~\ref{alg::solver}). Therefore, regarding the motivation discussed at the beginning of this Section~\ref{sect:main}, the proposed algorithm not only achieves efficient convergence as the Newton-type algorithm but also achieves an efficient active set detection mechanism via the concept of explicit MPC (i.e., mpQPs). Hence, in comparison with the standard SQP, the detection of active sets via mpQPs makes the proposed algorithm advantageous. Additionally, such a benefit does not significantly increase the iteration cost, which retains a low absolute computational time in practice. This is not the case in other SQP-style extensions, such as the ALADIN method, and we postpone the detailed comparison with ALADIN to Section~\ref{sect:compare}.}

\subsection{Dual Interpretation}\label{sect:dual}
\change{In this part, we would show a different but more intuitive view of the proposed algorithm. The expansion of the first-order information $g_k$ used in~\eqref{eq::coQP} gives
\begin{align*}
    g_k = \left(\mathrm{diag}(R_{k-1},Q_k)+\rho\mathbf{I}_{n_x+n_u}\right)\xi_k+[
    r_{k-1}^\top;q_k^\top]^\top+P_\xi^k\kappa_k.
\end{align*}
Moreover, recall that $P_\xi$ is the parameter of the inequality constraints~\eqref{eqn:cons_combined}, and $\kappa_k$ are the corresponding dual variables~\eqref{eq::probNeq}, which are generated by the mpQPs solutions directly. By inspecting the objective function in~\eqref{eq::coQP}, the quadratic penalty term $\mu\lVert s_k\rVert^2$ and $g_k$ together recover an augmented Lagrangian defined at $\xi$, where the active inequality constraints are dualized. This observation leads to a dual interpretation of the proposed algorithm.} \change{In the local problems~\eqref{eq::deQP} (i.e., step 1) in Algorithm~\ref{alg::solver}), the system dynamics are dualized with fixed dual variables (i.e., $\{\lambda_k\}$) based on the proximal-point Lagrangian. The dual variables to the inequality constraints (i.e., $\{\kappa_k\}$) are thereby updated. Accordingly, the coupled QP problem~\eqref{eq::coQP} (i.e., step 4) in Algorithm~\ref{alg::solver}) updates the dynamics dual with the dual variables to the inequality constraints fixed. Hence, the proposed algorithm can be viewed as an alternating direction method in the dual space. Via the scope of this dual interpretation, the coupled QP~\eqref{eq::coQP} is not a relaxed problem, as the augmented Lagrangian is an exact penalty function~\cite[Chapter 17]{Nocedal2006}.} \newchange{More specifically, due to the local convergence of the dual variables by Theorem~\ref{thm:convergence}, the augmented Lagrangian converges to an exact penalty.}

\change{With this dual interpretation at hand, we are ready to elaborate on the reasoning behind the use of proximal-point Lagrangian. Firstly, the dual variables model first-order local properties~\cite{rockafellar1974augmented}, and an aggressive update should therefore be avoided due to such locality. The proximal term (i.e., $\frac{\rho}{2}\lVert \xi_k-\overline{\xi}_k\rVert^2$ in the local steps~\eqref{eq::deQP}) realize this goal. This is important, especially when a good estimate of dual variables is not yet available. Secondly, as dual variables encompass first-order information, linearization is therefore needed. More specifically, the dual to the dynamics is updated to $\overline{\xi}$ in the coupled QP step. Hence the proximal-point Lagrangian is linearized around $\overline{\xi}$ in~\eqref{eq::deQP}. It is worth noting that the design logic similar to the aforementioned one also appears in other nonconvex primal alternating direction methods, such as~\cite{bolte2014proximal}. To the best of our knowledge, the proposed algorithm is the first algorithm to bring this idea to the dual space.}

\begin{remark}
\newchange{The penalty parameters $\rho$ in~\eqref{eq::deQP} and $\mu$ in~\eqref{eq::coQP} play a crucial role in determining the performance of the algorithm. A larger penalty typically leads to better convergence performance~\cite{rockafellar1974augmented}, but also results in more iterations and longer computational times. Although the convergence aspect of penalty weight has been extensively studied in the literature~\cite{rockafellar1974augmented,bertsekas1976multiplier}, it remains unclear how to select a penalty that balances convergence performance and the absolute solution time. We leave this for future investigation.}
\end{remark}

\subsection{Comparison with Related Work}\label{sect:compare}
\change{In this part, we will compare the proposed scheme with other related results, particularly the augmented Lagrangian based alternating direction inexact Newton (ALADIN) method. The ALADIN method is also an extension of the SQP algorithm, but it can only handle linear coupling. Hence, auxiliary variables that duplicate the states are introduced to handle the bilinear dynamics. More specifically, ALADIN reformulates the bilinear MPC problem~\eqref{eqn:prob_glb} to the following equivalent problem:}
\change{\begin{subequations}\label{eqn:aladin_prob_glb}
\begin{align*}
\min_{x_0,x,u}\;\;\sum_{k=0}^{N-1} \ell_k(x_k,u_k)+\ell_N(x_N)
\end{align*}
subject to:
\begin{align}
& x_0 = x(t),\notag\\
&\forall\, k\in\{0,1,...,N-1\},\notag\\
&z_{k+1}=Ax_k + Bu_k + \sum_{i=1}^{n_u}C_{i} x_k[u_{k}]_i+B_ww_k\notag\\
&x_{k+1} = z_{k+1}\;\; \mid \tilde{\lambda}_k,\label{eqn:aladin_cpl}\\
& z_{k+1}\in \mathcal{X},\; x_k\in \mathcal{X},\;u_k\in \mathcal{U}\notag
\end{align}
\end{subequations}
where auxilary variables $z_k$ duplicate the states $x_k$, and a linear coupling constraint is thereby introduced in~\eqref{eqn:aladin_cpl}. This is the standard horizon splitting scheme used in other nonlinear MPC algorithms~\cite{deng2019parallel,jiang2019time,laine2019parallelizing,shin2019parallel}, where inputs and states of the same time step are grouped. This leads to the first advantage of this paper's proposed splitting scheme: no auxiliary variables are introduced, so the problem size remains unchanged. This benefit also leads to a limitation of the proposed algorithm: the proposed splitting scheme requires that the stage cost and the constraints are decoupled between states and inputs. How to overcome this limitation is left for future research.}

\change{In the ALADIN algorithm,  
following $N$ nonconvex subproblems 
$\{\mathcal{P}_k\}_{k=0}^{N-1}$ can be solved in parallel:
\begin{subequations}
\begin{align*}
\forall k\in \{0,1,\dots,N-2\}
\end{align*}
\begin{align*}
\mathcal{P}_k:=
\min_{\substack{x_k,u_k\\z_{k+1}}}& \ell_k(x_k,u_k)+\begin{bmatrix}
\tilde{\lambda}_{k-1}\\-\tilde{\lambda}_k
\end{bmatrix}^\top\begin{bmatrix}
x_k\\ z_{k+1}\end{bmatrix}\\&+ \frac{\rho}{2}\left\lVert\begin{bmatrix}
x_k\\z_{k+1}\end{bmatrix}-\begin{bmatrix}
\overline{x}_k\\\overline{z}_{k+1}
\end{bmatrix}\right\rVert^2
\end{align*}
\text{subject to:}
\begin{align*}
&\; z_{k+1}=Ax_k + Bu_k + \sum_{i=1}^{n_u}C_{i} x_k[u_{k}]_i+B_ww_k\\
& z_{k+1}\in \mathcal{X},\; x_k\in \mathcal{X},\;u_k\in \mathcal{U}
\end{align*}
\begin{align*}
\mathcal{P}_{N-1}:=
\min_{\substack{x_{N-1},u_{N-1}\\z_N}}& \ell_k(x_{N-1},u_{N-1})+\ell(z_N)+\tilde{\lambda}_{N-1}^\top
x_{N-1}\\&+ \frac{\rho}{2}\lVert
x_k-\overline{x}_k\rVert^2
\end{align*}
\text{subject to:}
\begin{align*}
&\; z_{N}=Ax_{N-1}+ Bu_{N-1} + \sum_{i=1}^{n_u}C_{i} x_k[u_{k}]_{N-1}+B_ww_{N-1}\\
& z_N\in \mathcal{X},\; x_{N-1}\in \mathcal{X},\;u_{N-1}\in \mathcal{U}
\end{align*}
\end{subequations}
}

\change{After the parallel iteration, the ALADIN algorithm applies a relaxed SQP-style step to the reformulated problem~\eqref{eqn:aladin_prob_glb} in order to update the coupling dual variables $\{\tilde{\lambda}_k\}$. The differences between the ALADIN and the proposed scheme now become clear:
\begin{itemize}
    \item The proposed scheme only needs to solve a convex QP, whose solutions can be precomputed offline via mpQPs. Instead, the ALADIN algorithm has to solve a nonconvex problem online. Even though these nonconvex subproblems can be computed in parallel, the resulting computational cost per iteration is still significantly higher than the proposed scheme.
    \item The proposed scheme directly handles the nonlinear coupling (i.e., the bilinear dynamics) and therefore does not need to introduce auxiliary variables to duplicate the states. As a result, the SQP step used in the proposed scheme solves a smaller problem than the one solved in the ALADIN.
\end{itemize}}

\change{Due to the use of an SQP-style update and the use of augmented Lagrangian methods~\cite[Section~4]{houska2016augmented}, the proposed algorithm and the ALADIN algorithm have a certain similarity. However, their focus during the algorithm design is different. ALADIN focuses more on allocating the computational complexity, while the proposed algorithm aims at efficient iteration with a good active detection scheme. That is why the ALADIN tends to handle the non-convexity directly, as this can be handled by different computational nodes. On the contrary, the proposed algorithm is customized to bilinear MPC to have a lower computational cost per iteration and fewer decision variables. In summary, even though both the ALADIN and the proposed scheme can be viewed as extensions to the SQP algorithm, the ALADIN is more tailored for distributed computation. The proposed scheme is instead tailored for efficient online computation. Finally, we wrap up this section by summarizing the benefits of the proposed scheme as follows:
\begin{itemize}
\item It brings the efficiency of explicit MPC into an NMPC setup. Integrating the explicit mpQP solution provides two benefits: it returns an accurate primal solution when good estimates of the dual variables $\{\lambda_k\}$ are given, and it significantly improves the real-time efficiency by providing the active set estimation.
\item It retains the SQP structure. This not only preserves the convergence rate of the SQP algorithm but also makes the proposed algorithm compatible with any existing acceleration strategy developed for real-time SQP, such as warm-start.
\item It enjoys high computational efficiency even without parallelization. With proper implementation, this efficiency may be further improved by parallelization for some processors.
\end{itemize}
}

\begin{remark}
\change{It is worth mentioning that the proposed scheme reduces to an ALADIN algorithm when the dynamics are linear (i.e., $C_k=0$). In this case, the resulting algorithm is similar to the one studied in~\cite{jiang2019time}, where global convergence is also guaranteed~\cite{houska2017convex}.}
\end{remark}
\begin{remark}
\change{The alternating direction method of multipliers (ADMM) can be used to solve a bi-convex optimization problem~\cite[Chapter 9.2]{boyd2011distributed}. With the standard formulation given in~\cite[Chapter 9.2]{boyd2011distributed}, the ADMM algorithm needs to solve a nonconvex QP problem in each iteration, which is proved to be NP-hard even for the calculation of a local minimizer~\cite{contesse1980caracterisation,forsgren1991identification}. The resulting computational cost per iteration is significantly higher, and such ADMM formulation is, therefore not suitable for our comparison. Suppose the proposed interlacing horizon splitting scheme is applied instead. In that case, the resulting ADMM algorithm gets rid of the solution of a nonconvex QP, which is also one contribution of this work. However, the ADMM algorithm is still not suitable for comparison. On the one hand, a convergence guarantee exists only when there is no state constraint, which is undesirable in MPC applications. Based on our test on the numerical \newchange{example} given in the following Section~\ref{sect:num_build}, we did not observe the convergence of the ADMM after 3000 iterations (equivalently 1 minute in absolute time). On the other hand, the bilinear dynamics are squared in the augmented Lagrangian. The resulting problem is no longer an mpQP, and cannot be precomputed offline. As a result, multiple inequality constraint QPs are required to be solved in each iteration, leading to a much higher computational cost. Finally, even though we did not observe convergence in our numerical study, if it happens to converge for some specific cases, the convergence rate of a nonconvex ADMM algorithm is at most sublinear~\cite{wang2019global}. Therefore, it requires more iterations and accordingly, more computational time to converge.}
\end{remark}

\section{Implementation Details}
\label{sec::impl}
This section elaborates on the implementation details of Algorithm~\ref{alg::solver} with a particular emphasis on run-time aspects and a limited memory requirement. Here, the implementation of Steps 3) and 5) turns out to be straightforward, such that we focus on the implementation of Steps 1), 2), and 4).

\subsection{mpQP Subproblems}\label{sect:mpQP}
We summarize the local mpQPs~\eqref{eq::deQP} into a uniform form
\begin{equation}
\label{eq::dempQP}
\mathds{P}(\theta_k):\;\;\min_{\xi_k\in\Xi_k}\;\; \frac{1}{2}\xi_k^\top \mathcal{Q}_k\xi_k + \theta_k^\top \xi_k
\end{equation}
with parametric inputs $\theta_k\in\mathbb R^{n_x+n_u}$ and coefficient matrices $\mathcal{Q}_k = \text{diag}(R_{k-1},Q_{k}) +\rho \mathbf I_{ n_x+n_u}$ for all $k\in\mathbb Z_1^{N}$. Here, the first problem is omitted as its solution is fixed by $\xi_0=x(t)$ due to the initial state constraint enforced by $\Xi_0$. Based on the formulation of $\mathcal{L}_k^\rho$, we can work out the explicit form of $\theta_k$ as follows,
\begin{subequations}
\begin{align}\notag
\theta_k=&[r_{k-1}^\top;q_k^\top]^\top + \left(E_{k-1}+\mathrm{mat}(G_{k-1}\overline{\xi}_{k-1})\cdot S_k\right)^\top \lambda_{k-1}\\
&+\left[D_k+ (S_{k+1}\overline{\xi}_{k+1} \otimes \mathbf I_{n_x})^\top G_k\right]^\top \lambda_k- \rho \overline{\xi}_k \\\notag
\theta_N=& [r_{N-1}^\top;q_N^\top]^\top - \rho \overline{\xi}_N\\
&+ \left(E_{N-1}+\mathrm{mat}(G_{N-1}\overline{\xi}_{N-1})\cdot S_N\right)^\top \lambda_{N-1}.
\end{align}
\end{subequations}
Evaluating these parameters only requires matrix-vector multiplications such that the complexity is $\mathcal{O}(N\cdot(n_x+n_u)^2)$.
In this paper, we use the enumeration-based multi-parametric QP algorithm from~\cite{HJ15} for generating solution maps $\xi^\star_k:\mathbb R^{n_x+n_u}\to\mathbb R^{n_x+n_u}$ of~\eqref{eq::dempQP}. The complexity of pre-processing the small-scale QPs~\eqref{eq::dempQP} depends on the number of critical regions $N_{\text{R},k}$ over which the PWA optimizers $\xi_k^\star(\cdot)$ are defined~\cite{BemEtal:aut:02}.
Here, we assume that each parametric QP is post-processed, off-line, to obtain binary search trees~\cite{TJB03} in $\mathcal{O}(N_{\text{R},k}^2)$ time. Once the trees are constructed, they provide for a fast evaluation of the solution maps in~\eqref{eq::dempQP} in time that is logarithmic in the number of
regions, thus establishing the $\mathcal{O}(\sum_k \log_2(N_{R,k}))$ on-line computational bound. The memory requirements are directly proportional to the number of critical regions, with each region represented by a finite number of affine half-spaces. Finally, it is worth mentioning that if \change{$Q_i = Q_j,\;q_i=q_j,\;R_i=R_j,\;r_i = r_j,\;\forall\;i\neq j$, then the $i$-th mpQP subproblems~\eqref{eq::dempQP} is identical to the $j$-th one, and one mpQP solution can therefore serve for two subproblems. Identical subproblems happen in many MPC applications as the stage cost are usually fixed throughout the prediction horizon.}

\subsection{Sensitivities Evaluation}
Step~2) of Algorithm~\ref{alg::solver} evaluates the sensitivities $g_k$, $c_k$, $\hat P_\xi^k$ and $H$. As we consider the quadratic cost, the gradients $g_k$ can be easily evaluated with analytical form. Moreover, the primal feasibility residual $c_k$ and active Jacobian $\hat P_\xi^k$ are also straightforward. Therefore, we focus on the computation of the Hessian matrix $H$ in this subsection. 

As we used an interlacing horizon splitting scheme, the exact Hessian $\nabla_{\xi\xi} L^0(\xi,\lambda,\xi)$ is not block diagonal with respect to each $y_k$ but the banded block diagonal. However, as the off-diagonal blocks only involve the bilinear dynamics, we can work out each block analytically as follows:
\[
\nabla_{\xi\xi} L^0(y,\lambda,y) = 
\begin{bmatrix}
\mathcal{Q}_0&S_{0,1}&&&\\
S_{1,0}&\mathcal{Q}_1&S_{1,2}&\\
&\ddots&\ddots&\ddots&\\
& & S_{N-1,N}&\mathcal Q_N
\end{bmatrix}
\]
with blocks
\[
\begin{aligned}
&S_{0,1}=S_{1,0}^\top= [
C_1^\top \lambda_0,..., C_{n_u}^\top \lambda_0,\mathbf{0}_{n_x\times n_x}
]\in\mathbb R^{n_x\times (n_u+n_x)}\\
&S_{k,k+1}=S_{k,k+1}^\top \\
=&\begin{bmatrix}
C_1^\top \lambda_{k}&...& C_{n_u}^\top \lambda_k&\mathbf{0}_{n_x\times n_x}\\
\mathbf{0}_{n_u}&...,&\mathbf{0}_{n_u}&\mathbf{0}_{n_u\times n_x}
\end{bmatrix}\in\mathbb R^{(n_x+n_u)\times (n_x+n_u)}
\end{aligned}
\]
for all $k\in\mathbb Z_1^{N-1}$. It is clear that evaluating the exact Hessian is equivalent to evaluating $C_i^\top\lambda_k$ for all $i\in\mathbb Z_1^{n_u}$ and $k\in\mathbb Z_0^{N-1}$. Therefore, its computational complexity is only $\mathcal{O}(Nn_un_x^2)$. 
In practice, some heuristics can be adopted to achieve better numerical robustness on the convergence performance of Algorithm~\ref{alg::solver} such as enforcing $H\approx \nabla_{\xi\xi} \mathcal L^0\succ 0$ by adding a regularization term, i.e., $H = \nabla_{\xi\xi} \mathcal L^0+ \sigma \mathbf I$ with $\sigma\geq 0$~\cite{verschueren2017sparsity}.

\subsection{Coupled QP}
The coupled QP~\eqref{eq::coQP} has no inequality constraints such that solving~\eqref{eq::coQP} is equivalent to solving linear equations defined by the KKT system:
\begin{equation}
\label{eq::KKT}
\underbrace{\begin{bmatrix}
H+\rho \hat P_\xi^\top \hat P_\xi & J^\top\\[0.12cm]
J 
\end{bmatrix}}_{\mathcal H}
\underbrace{
\begin{bmatrix}
\Delta \xi \\[0.12cm]
\lambda^\text{QP}
\end{bmatrix}}_{w}=
\begin{bmatrix}
-g \\[0.12cm] -c
\end{bmatrix}
\end{equation}
with
\[
J = \begin{bmatrix}
\tilde D_0 & \tilde E_0 & & & \\[0.12cm]
&\tilde D_1 & \tilde E_1 & &\\[0.12cm]
&&\ddots &\ddots &\\[0.12cm]
&&&\tilde D_N & \tilde E_N
\end{bmatrix},
\begin{aligned}
\hat P_\xi &= \text{diag}(\hat P_\xi^1,...,\hat P_\xi^N)\\[0.12cm]
g & = [g_0^\top,g_1^\top,...,g_N^\top]^\top\\[0.12cm]
c & = [c_0^\top,c_1^\top,...,c_N^\top]^\top 
\end{aligned}
\]
and for all $k\in\mathcal Z_0^N$,
\[
\begin{aligned}
\widetilde D_k =& D_k + (S_{k+1}\overline{\xi}_{k+1} \otimes \mathbf I_{n_x})^\top G_k,\\
\widetilde E_k =&E_k+ \mathrm{mat}(G_{k}z_{k})\cdot S_{k+1}.
\end{aligned}
\]
If we rearrange the KKT matrix $\mathcal H$ by resorting $w$ as 
\[
(\Delta \xi_0,\;\lambda_0^\text{QP},\;\Delta \xi_1, \;\lambda_1^\text{QP},...,\Delta \xi_{N-1},\;\lambda_{N-1}^\text{QP},\;\Delta \xi_N),
\]
a tri-blocked-diagonal sparsity pattern appears in the KKT matrix $\mathcal H$, such that the Schur complement based back-forward sweeps can be used to solve the linear equation efficiently. To better illustrate this idea, we consider $N=2$ such that the resulting rearranged KKT system is 
\[
\begin{bmatrix}
\begin{array}{cccc|c}
\mathcal{Q}_0 &\widetilde D_0^\top & S_{0,1}&&\\[0.12cm]
\widetilde D_0&&\widetilde E_0&&\\[0.12cm]
S_{1,0}&\widetilde E_0^\top & \widetilde{\mathcal{Q}}_1&\widetilde D_1^\top& S_{1,2}\\[0.12cm]
&&\widetilde{D}_1&& \widetilde{E}_1\\[0.12cm]\hline
&&S_{2,1}&\widetilde{E}_1^\top&\widetilde{\mathcal{Q}}_{2}
\end{array}
\end{bmatrix}
\begin{bmatrix}
\Delta \xi_0\\[0.12cm] 
\lambda_0^\mathrm{QP}\\[0.12cm] 
\Delta \xi_1\\[0.12cm] 
\lambda_1^\mathrm{QP}\\[0.12cm]
\Delta \xi_2
\end{bmatrix}
=
\begin{bmatrix}
-g_0\\[0.12cm] 
-c_0\\[0.12cm] 
-g_1\\[0.12cm] 
-c_1\\[0.12cm]
-g_2
\end{bmatrix}
\]
with 
$\widetilde{\mathcal{Q}}_k =\text{diag}(R_{k-1}, Q_k)+\mu(\hat{P}_\xi^k)^\top \hat P_\xi^k $.
We start the backward sweep by considering the whole KKT matrix as a 2x2 block matrix. Then, applying the Schur complement with respect to the lower left block $\widetilde{\mathcal{Q}}_2$ yields a reduced KKT matrix
\[
\begin{bmatrix}
\begin{array}{cc|cc}
\mathcal{Q}_0 &\widetilde D_0^\top & S_{0,1}&\\[0.12cm]
\widetilde D_0&&\widetilde E_0&\\[0.12cm]\hline
S_{1,0}&\widetilde E_0^\top & \widetilde{\mathcal{Q}}_1+S_{1,2}\widetilde{\mathcal{Q}}_2^{-1}S_{2,1}&\widetilde D_1^\top +S_{1,2}\widetilde{\mathcal{Q}}_2^{-1}\widetilde E_1^\top\\[0.12cm]
&&\widetilde{D}_1+\widetilde E_1\widetilde{\mathcal{Q}}_2^{-1}S_{2,1}& \widetilde E_1\widetilde{\mathcal{Q}}_2^{-1}\widetilde E_1^\top
\end{array}
\end{bmatrix}.
\]
Applying the Schur complement once more results in a reduced KKT system with respect to only $(\Delta \xi_0,\lambda_0)$ such that the substitution of the initial condition $\Delta \xi_0 = 0$ can enable a forward sweep to recover the primal-dual solution $(\Delta \xi,\lambda)$. This method has been shown that it is equivalent to the Riccati recursion in dealing with LQR problems~\cite{frison2016algorithms}. As the update of the right-hand side of the KKT system only requires matrix-vector multiplication, we observe that the computational complexity of this linear solver is dominated by the matrix update (i.e., computation of the Schur complement), which is $\mathcal{O}(N(n_x+n_u)^3)$.

\section{Numerical Results}\label{sect:num}
This section studies the proposed algorithm on two bilinear system examples. The proposed algorithm is first compared against other state-of-art solvers on a building control problem running on a laptop computer. The algorithm is then implemented in an embedded microcontroller for speed control of a DC motor.
The binary search tree of the mpQP solutions used in the proposed algorithm is generated by the multi-parametric toolbox (\texttt{MPT 3.0})~\cite{kvasnica2004multi}. 

\subsection{Bilinear Building Control}\label{sect:num_build}
\change{In this part, the proposed algorithm is compared with an efficient optimal control solver \texttt{acados}~\cite{Verschueren2021} and the ALADIN algorithm, which is implemented by \texttt{ALADIN}-$\alpha$ toolbox~\cite{engelmann2022aladin}. The code generation in \texttt{acados} is based on the SQP method with exact Hessians and without/with condensing. All the algorithms use the mirror method to regularize the indefinite QP problem~\cite{verschueren2017sparsity}. It is worth mentioning that \texttt{acados} is highly optimized for MPC, whose linear algebra subroutine \texttt{BLASFEO}~\cite{frison2018blasfeo} and QP solver~\cite{frison2020hpipm} exploit the structure in MPC. On top of that, a sophisticated, active set detection scheme by exact penalty function is implemented in \texttt{acados}~\cite[Chapter 2.3]{Verschueren2021}. Hence, this comparison can demonstrate the performance of the proposed algorithm.}

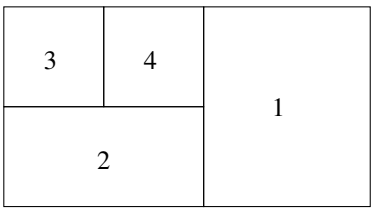
\begin{figure}[!htbp]
    \centering
    \begin{tikzpicture}
    \draw [draw=black] (0,0) rectangle (.3\linewidth,.15\linewidth);
    \draw [draw=black] (0,.15\linewidth) rectangle (.15\linewidth,.3\linewidth);
    \draw [draw=black] (.15\linewidth,.15\linewidth) rectangle (.3\linewidth,.3\linewidth);
    \draw [draw=black] (.3\linewidth,0) rectangle (.55\linewidth,.3\linewidth);
    \node at (0.15\linewidth,.07\linewidth) {2};
    \node at (0.07\linewidth,.22\linewidth) {3};
    \node at (0.22\linewidth,.22\linewidth) {4};
    \node at (0.412\linewidth,.15\linewidth) {1};
    \end{tikzpicture}
    \caption{Schematic diagram of the multi-zone building}
    \label{fig:multizone_scheme}
\end{figure}

 \change{We considered a multi-zone building model reported in~\cite{belic2021detailed} with room indices shown in  Figure.~\ref{fig:multizone_scheme}. Due to the space limit, the parameters of the model (i.e., $A,\;B,\;B_w,\;C,$ matrices) are included in the supplementary material on GitHub. In this multi-zone building (Figure~\ref{fig:multizone_scheme}), room 2 is the corridor linking a large warehouse (room 1) and two offices (rooms 3 and 4). An independent HVAC system controls the indoor temperature of room 1, while another HVAC controls the temperature of all other rooms. The corresponding control inputs ($u\in\mathbb{R}^2$) are the valve positions in the air handling unit, where the heat transfer between the air and the hot water flowing in the heating coil results in the bilinear term in the system dynamics. As a result, the control inputs can manipulate the supply air temperature in a nonlinear way, which accordingly controls the indoor temperatures. In summary, this is a 15-dimensional model (i.e., $x\in\mathbb{R}^{15}$) with two-dimensional control inputs, the states include the indoor temperature, wall temperature between two different rooms, wall temperature that stands between a specific room, and outdoor, and supply air temperature control. Process noises are outdoor temperature and solar radiation (i.e., $w\in\mathbb{R}^2$). In building control, a common practice is to apply certainty equivalence control~\cite{oldewurtel2013stochastic}, which uses weather forecast as the nominal disturbance in the MPC formulation. Meanwhile, the building evolves under the actual weather condition that is similar but not identical to the weather forecast. Real-world weather data is used for this numerical study.}
 
\change{Using this approach, 100 Monte-Carlo tests were conducted with recorded weather from \href{tomorrow.io}{tomorrow.io}~\cite{tomorrow} in winter (Fig.\ref{fig:building}(a)-(b) plot one sampled weather condition). The weather forecast used in the MPC problem is the recorded weather perturbed by zero-mean random noise, while the simulation uses the recorded weather (i.e., the weather forecast curves in Figure~\ref{fig:building}~(a)-(b)). The prediction horizon is set to 8 with an objective of minimizing energy consumption, whose loss function is
\begin{align*}
\ell_k(x_k,u_k) =  u_k^2\;.
\end{align*}
To ensure occupant comfort, the indoor temperature is bounded within $[22,24]^\circ C$. The control input (i.e., fractional valve position) is bounded within $[0,1]$.}
\begin{remark}
It is possible to define an objective as $\ell_k(x_k,u_k) =  |u_k|$. The resulting local problem can be reformulated as a linear program and, thus, also an mpQP. We use a quadratic loss function here to avoid unnecessary confusion.
\end{remark}

\change{For this numerical test, all solvers use the same initialization in the first iteration and apply the same warm-start strategy to generate the initialization for the following iterations. In particular, the shifted solution from the last iteration is used to warm-start. The computational time is the sum of the CPU time returned by the solver. The results in this subsection are generated by a laptop with an Intel i7-11800H 16-core processor and 32 GB memory. Meanwhile, as Step 1) in the proposed algorithm~\ref{alg::solver} is a convex QP, the solution time without using mpQPs is also investigated. In particular, the mpQP solution in this example has 729 critical regions, and the resulting binary search tree is of depth 13 (Section~\ref{sect:mpQP}). The parallelization of the proposed algorithm is done by using \texttt{OpenMP}, while the parallelization in ALADIN relies on the parallel computing toolbox from \texttt{Matlab}. The statistics of the solution time are summarized in Table~\ref{tab:benchmark}, where the maximal solution time indicates the solution time when \newchange{a} good initialization is not available (i.e., cold-start). \newchange{In contrast}, the mean solution time reflects the averaged performance when a good initialization is available.}

\change{Above all, Table~\ref{tab:benchmark} shows that the ALADIN method is not desirable for fast MPC applications, the need to solve multiple nonconvex problems significantly slows down its speed (Section~\ref{sect:compare}). We only report the parallelized solution time for ALADIN, and the non-parallelized solution time is at least three times slower. Regarding the proposed algorithm, the overhead caused by parallelization pays off only when mpQP solutions are not used. In this case, step 1) in the proposed algorithm requires solving a QP whose computational cost is significantly higher than calling the mpQP solution. Thus, it is easier to improve performance by parallelization in this case. On the contrary, as calling mpQP solution is already computationally highly efficient, improving performance by parallelization may require more involved code design, such as caching. We believe that is the reason why the use of \texttt{OpenMP} does not accelerate the mpQP-based implementation in this case. Note that this observation does not negate the benefit of parallelization. On the one hand, the efficiency of parallelization depends on the computing unit and the compiler. The use of \texttt{OpenMP} and a general purpose \texttt{Intel} CPU in this numerical example may not be the most efficient implementation. On the other hand, constructing the mpQP solutions may not be computationally affordable for large-scale systems. If solving convex QPs  online is needed, then performance improvement is easy to achieve by parallelization, which is also justified by this numerical test.}

\change{\newchange{The comparison with \texttt{acodos} is shown in Table~\ref{tab:benchmark}}. When the proposed scheme uses mpQPs solution without parallelization, \newchange{its maximal solution time} is on average 71\% faster than that of \texttt{acados} without condensing. Regarding the mean solution time, even though the proposed scheme is faster than \texttt{acados} without condensing by 17\%, but it is 48\% slower than the mean solution of \texttt{acados} with condensing. Note that both the QP solver and the linear algebra routine in \texttt{acados} is highly optimized for NMPC; the results in Table~\ref{tab:benchmark} show that a tailored SQP solver is highly efficient when a good initialization is available. As Step 4) in the proposed scheme is similar to an SQP iteration, the proposed algorithm also shows comparable performance to a tailored SQP solver for warm-started iterations. This computational efficiency aligns with our discussion in Section~\ref{sect:main}. On top of this benefit, the proposed scheme performs much better when a good initialization is unavailable. This justifies the use of mpQP solutions, which improves the detection of the active sets. In summary, this numerical study proves the efficiency and efficacy of the proposed algorithm, and it also suggests a further possibility of performance improvement by a more sophisticated combination of the proposed scheme and a tailored SQP solver; we leave this for a future study.}

\change{Besides the observation given in Table~\ref{tab:benchmark}, we also observe that the proposed algorithm is more robust to the choice of initialization strategy. More specifically, if the initialization is only partially warm-started by setting the predictive input sequence to $\mathbf{0}$ (i.e., cold-start inputs but warm-start all the other variables), both \texttt{acados} and ALADIN will return NaN during the simulation for all the Monte-Carlo tests. On the contrary, the proposed scheme will always converge even when all the decision variables are initialized by $\mathbf{0}$. This observation aligns with the motivation of the proposed algorithm and justifies the benefit of Step 1) in the proposed algorithm. Meanwhile, this robustness might be beneficial in some applications. For example, set-point change in tracking control makes initialization more challenging. }

\begin{table*}[htbp!]
\renewcommand{\arraystretch}{1.7}
\centering
\caption{\change{Statistics of the solution time at different tolerance (the entries of the top two performers in each row are stressed by boldface black and boldface blue respectively)}}
\begin{tabular}{c|c|c|c|c|c|c|c|c}
\hline
     \multicolumn{2}{c|}{method} & \multicolumn{4}{c|}{Algorithm~\ref{alg::solver}} &\multicolumn{2}{c|}{\texttt{acados}}& \texttt{ALADIN}\\\hline
     \multirow{2}{*}{tol} & \multirow{2}{*}{\shortstack{sol time\\(ms)}} & \multicolumn{4}{c|}{parallel (mpQPs)}& \multicolumn{2}{c|}{condensing} & \multirow{2}{*}{parallel}\\\cline{3-8}
     &&\shortstack{yes\\(yes)}&\shortstack{no\\(yes)}&\shortstack{yes\\(no)}&\shortstack{no\\(no)}&yes&no&\\\hline
     \multirow{2}{*}{$10^{-4}$}& max & \textbf{\color{black}8.404}& \textbf{6.304} & 8.926 & 12.272 & 19.858&10.711 & 2553\\\cline{2-9}
     &mean& 0.5280  & \textbf{\color{black}0.4835} & 0.6583 & 0.8751 & \textbf{0.3671} & 0.7931 &897.4\\\hline 
     \multirow{2}{*}{$10^{-5}$}& max & \textbf{\color{black}8.838} & \textbf{6.629} & 9.371& 12.905&21.755 & 11.804&2859\\\cline{2-9}
     &mean& 0.6864 & \textbf{\color{black}0.6286} & 0.8558  & 1.137& \textbf{0.4020}& 0.8687& 953.2\\\hline 
     \multirow{2}{*}{$10^{-6}$}& max& \textbf{\color{black}8.968} & \textbf{6.753} & 9.593& 13.510 &22.532 & 12.251 & 3369\\\cline{2-9}
     &mean& 0.7814 & \textbf{\color{black}0.7156} & 0.9743 & 1.295 & \textbf{0.4163} & 0.8995 &1053\\\hline 
\end{tabular}
\label{tab:benchmark}
\end{table*}

\input{building_plot}

\change{Last but not least, the property that the proposed algorithm is feasible even with an infeasible initial state is useful in practice, which is typically the case in building control. Due to the uncertain occupant behavior, such as opening the window, the indoor climate can be significantly perturbed, resulting in an infeasible initial state for the MPC problem. Consider a case where the occupant opens the window to bring in the fresh air when he first arrives in room 1 at  10:00 A.M., this move causes a sudden drop in indoor temperature as shown in Fig.~\ref{fig:building} (c). Such sudden temperature drop causes infeasibility, which leads to the failures of the \texttt{acados} solver. However, the proposed algorithm can still give reasonable control inputs and quickly recovers the indoor temperature to a comfortable level.}

\subsection{Bilinear DC Motor Control with a C2000 Microcontroller}
\begin{figure}[htbp!]
\begin{subfigure}[b]{\linewidth}
 \centering
 \includegraphics[width=\linewidth]{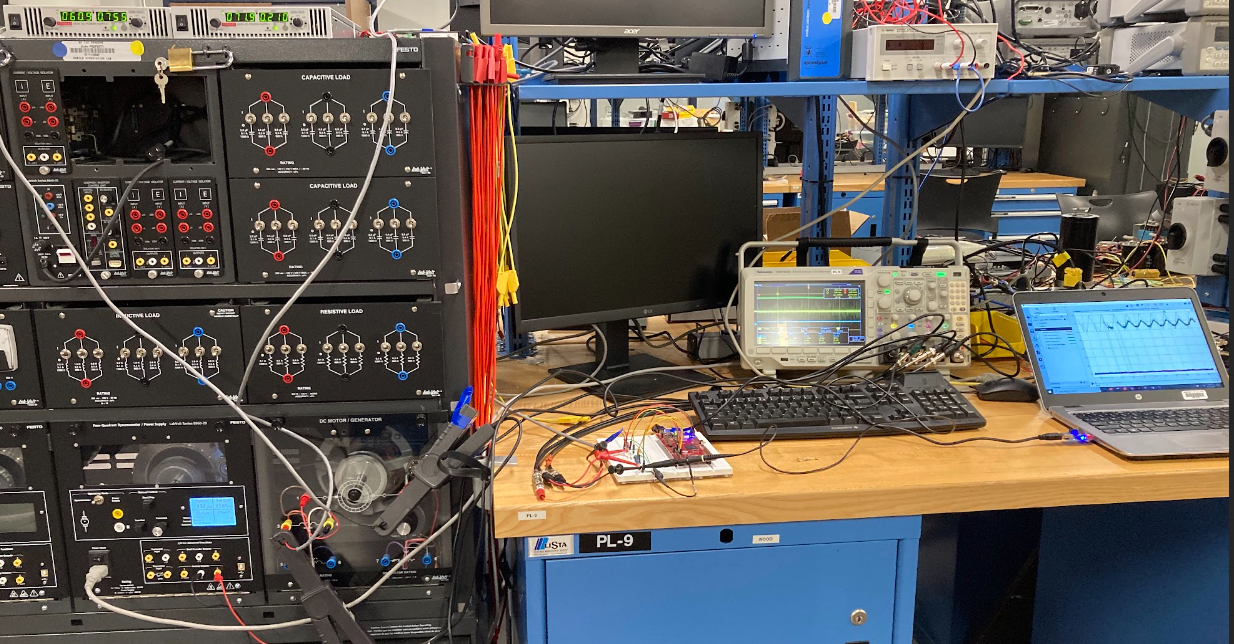}
 \caption{Overall Test setup}
\label{fig:hardware_setup}
 \end{subfigure}
\begin{subfigure}[b]{\linewidth}
 \centering
 \includegraphics[width=\linewidth]
 {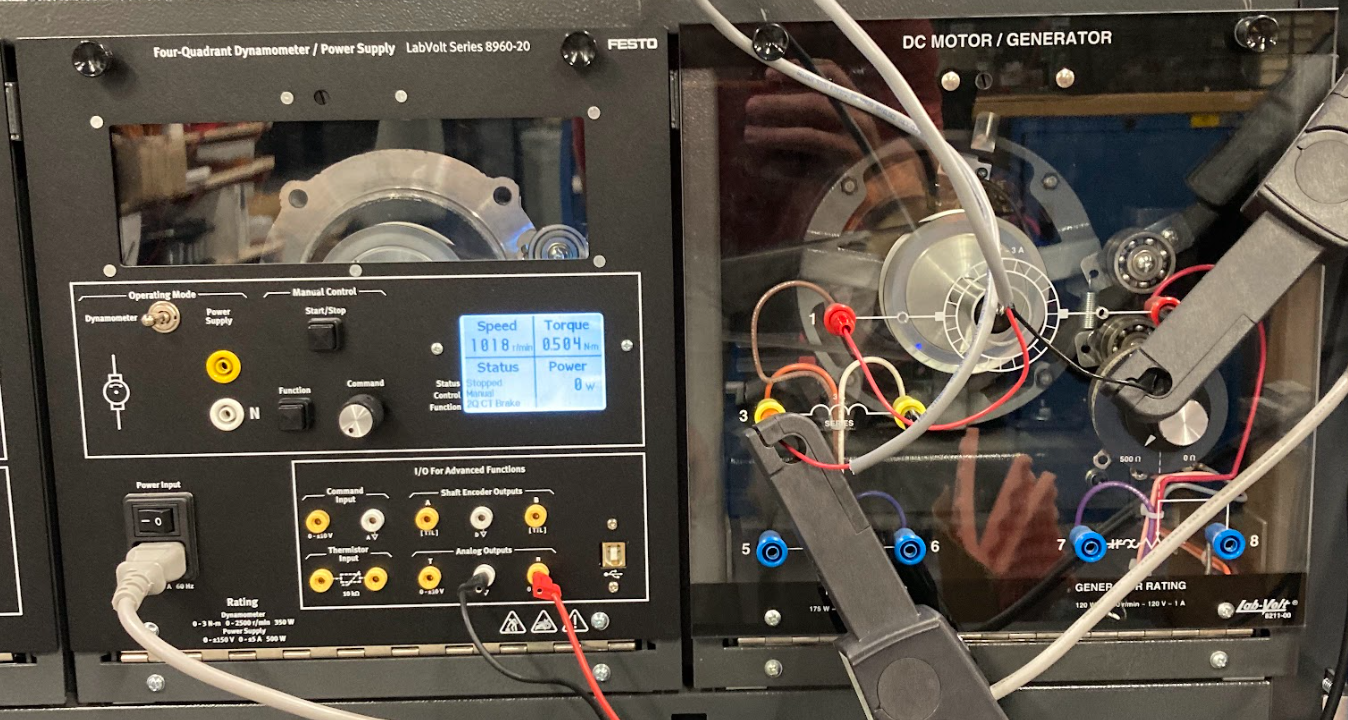}
 \caption{Dynamometer (left) and Motor (right)}
\label{fig:hardware_motor_dyno}
 \end{subfigure}
\begin{subfigure}[b]{\linewidth}
 \centering
 \includegraphics[width=\linewidth]
{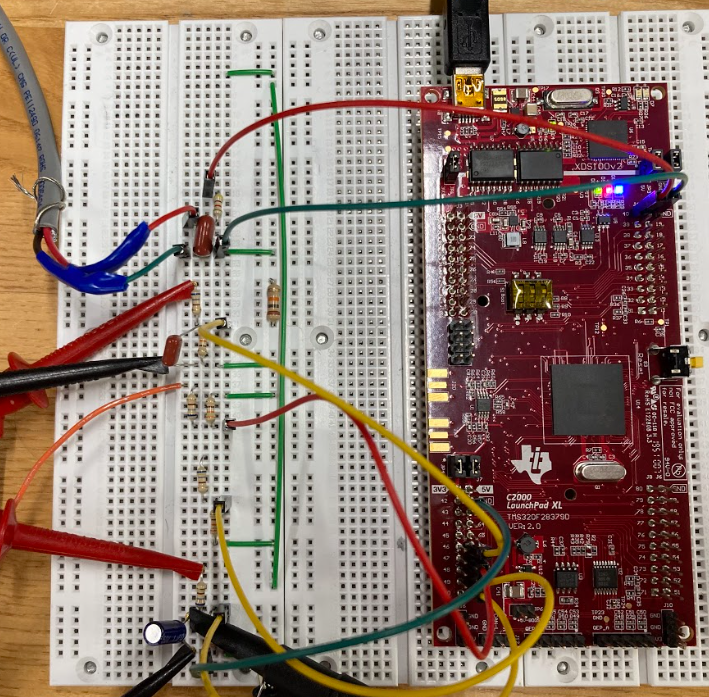}
 \caption{TI LaunchPad XL and analog I/O filtering}
\label{fig:C2000_board}
 \end{subfigure}
\caption{Hardware testing of the proposed MPC algorithm with a real motor controlled by a TI C2000 Delfino 28379D microcontroller. The overall test setup is visible in (a) with the host computer on the right, the power supplies for the armature and field windings in the top left, the motor/dynamometer setup in the bottom left, and the controller in the front left corner of the bench. A detailed view of the motor is shown in (b), with clamp-on current sensors for the armature and field currents. The controller and its associated analog input/output filters are shown in (c).}
\label{fig:motor_hardwareMPC}
\end{figure}

Next, the proposed algorithm is deployed on an embedded system, a Texas Instruments C2000 LaunchPad XL F28379D, to control the speed of a field-controlled DC motor. The dynamics of the field-controlled DC motor are bilinear,
\begin{align*}
\frac{dx_1}{dt} &= -\frac{R_a}{L_a}x_1-\frac{K_m}{L_a}x_2u+\frac{V_s}{L_a},\\
\frac{dx_2}{dt} &= -\frac{B}{J}x_2+\frac{K_m}{J}x_1u-\frac{T_e}{J},
\end{align*}
where states $x_1$, $x_2$ are, respectively, armature current and angular velocity, and the control input $u$ is field current. $V_s$ and $T_e$ define the external torque, respectively, which are chosen as 60 V and 0 Nm for this experiment. The remaining parameters are identified on a real field-controlled DC motor (Fig.~\ref{fig:hardware_motor_dyno}) as shown in Table~\ref{tab:motor_params}.

\begin{table}[htbp!]
\renewcommand{\arraystretch}{1.5}
\centering
\caption{Parameters of the Field Controlled DC Motor}
\begin{tabular}{c|c|c}
\hline
     Parameter & Variable & Value\\\hline
     Armature Resistance & $R_a$ & 10 $[\mathrm{ohm}]$ \\
     Armature Inductance & $L_a$ & 60 $[\mathrm{mH}]$ \\
     Motor Constant & $K_m$ & 0.2297 $[\mathrm{V~(A~rad/s)^{-1}}]$ \\
     Damping Ratio & B & 0.0024 $[\mathrm{Nm(rad/s)^{-1}}]$ \\
     Inertia & J & 0.008949 $[\mathrm{kg~m^2}]$ \\
     \hline
\end{tabular}
\label{tab:motor_params}
\end{table}

We first provide some background on the motor behavior to gain insight into the NMPC solutions. Typically, the armature current dynamics $x_1$ are much faster than the mechanical dynamics, so it is useful to consider the motor behavior after the current dynamics have decayed. The torque-speed curves of the motor are shown in Fig.~\ref{fig:current_curve}~(a) for various field currents. This is the electrical torque produced by the motor for the given speed and field current. If the mechanical torques (drag+external) match this torque at a given speed, it is an equilibrium point. For example, we may observe that the no-load speed for this motor (without drag) is 87 rad/s at the full 3~A field current. The typical operating region of this type of motor is at speeds higher than the full-field line, roughly $[80,180]\,\textrm{rad/s}$ for low torques. Operation below this speed is undesirable because the armature currents exceed the $3\,\mathrm{A}$ continuous thermal limits regardless of the field current selected. \change{This is shown in the current-speed curve (Fig.~\ref{fig:current_curve}~(b)), where armature current is plotted as a function of speed for different field currents (i.e., control input). Hence, the curves below the red dashed line in Figure~\ref{fig:current_curve}~(b) also show the set of desired operating points that allows long-term operation without overheating.}

The continuous dynamics are discretized by the Euler method with a sampling time of 10ms. The prediction horizon is set to 3\footnote{\change{The prediction horizon is set based on some recent results with \textbf{commercial} solver from \texttt{ODYS}~\cite{cimini2015online,cimini2020embedded}, where they use the same C2000-series hardware to deploy MPC on a synchronous machine. In their setup, the input constraints are neglected, the prediction horizon is two, and the linearized model is used instead of the nonlinear model.}} with a convergence tolerance at $10^{-4}$. The MPC controller conducts speed control, which tracks a reference speed $\omega_{\mathrm{ref}}$. This motor operates around [80,180]~rad/s, and for most reference torque/speed combinations within this range, there are two possible field current solutions as shown by overlapping lines in Fig.~\ref{fig:current_curve}~(a). The low field current (i.e., control input) solution results in a higher armature current, usually above the 3~A limit (Fig.~\ref{fig:current_curve}~(b)). Long-term operation on this equilibrium point will result in armature overheating even though it tracks the reference speed. However, to have an agile motor response, the armature current should be able to operate above 3 A for short intervals. Therefore, we do not enforce a constraint on the armature current, while the field current (i.e., control input) is bounded within [1,3]~A. 

\change{In this control setup, the desired operating point has an armature current lower than 3~A, which corresponds to the higher of two viable field currents (Fig~\ref{fig:current_curve}). A proper choice of the loss function can help the solver to converge to the desired operating point.} First of all, it is not desirable to use the speed regulating stage cost $\ell(x,u)= ([x_{k}]_2-\omega_{\textrm{ref}})^2$, as the solver tends to select the lower field current command which will overheat the armature. This is particularly the case in the presence of noise based on our observation of different hardware in the loop simulations. We suspect that this can be explained by the torque-speed curve (Fig.~\ref{fig:current_curve}~(a)). When operating with a low field current, the curves are relatively flat, and a slight change in the field current can lead to a rapid change in speed. \change{This implies that the solver can give better local convergence behavior in this region, so the solver tends to converge to this undesired operating point.} To avoid this issue and push the solution to the preferred operating point, we offer a reference armature current and field current, whose steady state solution has an explicit form by substituting $\omega_{\textrm{ref}}$ into the system dynamics. The stage cost is designed to 
$$\ell(x,u)= 20([x_{k}]_1-I_{\textrm{ref}})^2+([x_{k}]_2-\omega_{\textrm{ref}})^2+10(u-u_{\textrm{ref}})^2,$$ 
where $I_{\textrm{ref}}$ and $u_{\textrm{ref}}$ are reference armature current and reference field current.

\input{current_curve}

The experimental setup is shown in Fig.~\ref{fig:motor_hardwareMPC} \change{with the explanation given in its caption}. Two experiments are carried out on velocity tracking control, which both track a triangular reference speed that varies between 100 rad/s and 140 rad/s. In the first case, we only have a speed constraint within $[80,180]$ rad/s, while this constraint is tightened to $[110,180]$~rad/s in the second case. Thus, the speed lower bound is inactive in the first case but active and satisfied in the second case.

To verify the real-time controller performance, we first simulate these experiments using control hardware in the loop methods. Both the controller and a simulated motor run on the same C2000 microcontroller (Fig. \ref{fig:C2000_board}). The results are shown in Fig.~\ref{fig:motor_sim_result}, with the armature current disturbed by white noise to reproduce the switching noise encountered in real-world experiments. In simulations, the NMPC properly executes the speed control, and the speed constraint is satisfied in the second case as expected. 

\change{The hardware in-the-loop result above can already justify the efficiency of the proposed solver, a similar setup for performance proof is also used in the \textbf{commerical} product in~\cite{cimini2015online,cimini2020embedded}. But for the sake of completeness, we carry out the experiment on a real motor.} The hardware experimental results are shown in Fig.~\ref{fig:motor_hardwareMPC_result}. The measured signals are post-processed with a low-pass zero-phase  filter. In this experiment, the proposed algorithm successfully executes the control in real-time with a 10 ms MPC update rate. In particular, the maximum and average execution time of the proposed algorithm in this embedded system are 2.088 ms and 1.764 ms, respectively. Thus, the solver can run up to $500$ Hz, which is sufficiently fast regarding the 10 ms sampling time of the targeted system.
 
However, in our real-world experiments, the tracking performance is somewhat lacking. From Fig.~\ref{fig:motor_hardwareMPC_result}, we can observe that the NMPC tries to track the signal, and periodic triangular speed trajectories are recorded with noticeable tracking errors. It is noteworthy that the NMPC satisfies the lower speed limit in the second test, which justifies the constraints enforced by the NMPC. The reasons will be investigated in future work but may be due to poor estimates of the unmeasured drag torque, other parameter errors, or inaccurate delivery of the current command. \change{In summary, based on the hardware in the loop experiments and the real-world experiments, the efficiency and real-time capability of the proposed algorithm has been proven. Even though it is not the main focus of this work, we believe that there are still a few \newchange{improvements} to be carried out on our experiment platform to exploit the capability of the NMPC control fully, and we leave this for future investigation.}

\input{motor_plot_sim}
\input{motor_plot}

        
\section{Conclusion}
This paper proposes a novel proximal-point Lagrangian based nonconvex solver for bilinear model predictive control. The proposed algorithm combines the ideas of explicit MPC, horizon splitting, and real-time SQP algorithms and a novel horizon splitting scheme is proposed to enable this integration. The numerical efficiency of the proposed algorithm is validated by a building control simulation and an experiment on a real field-controlled DC motor with a TI C2000 LaunchPad XL F28379D microcontroller. Particularly, the latter experiment proves the real-time capability of the proposed algorithm, which successfully solves the NMPC problem in 1.764 ms on average.

\bibliographystyle{ieeetr}
\bibliography{ref}

\begin{IEEEbiography}[{\includegraphics[width=1in,height=1.25in,clip,keepaspectratio]{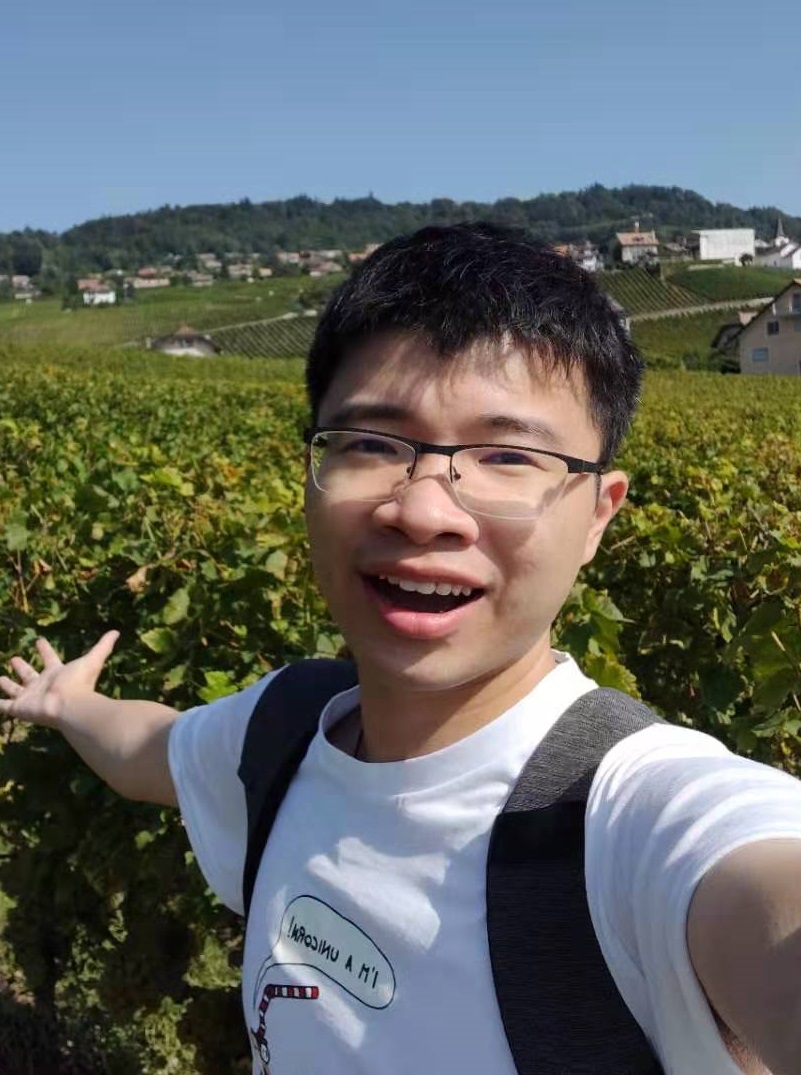}}]
{Yingzhao Lian} graduated with an MSc in Mechanical Engineering from EPFL in 2018. He specialized in optimization and learning theory. He jointly conducted his master’s thesis with the Automatic Control Lab, EPFL, and ABB Corporate Research, Baden, where he investigated the problem of data-driven model-based optimal control. In August 2018, he joined the Automatic Control Laboratory at EPFL as a Ph.D. student under the supervision of Professor Colin Jones.

\end{IEEEbiography}

\begin{IEEEbiography}[{\includegraphics[width=1in,height=1.25in,clip,keepaspectratio]{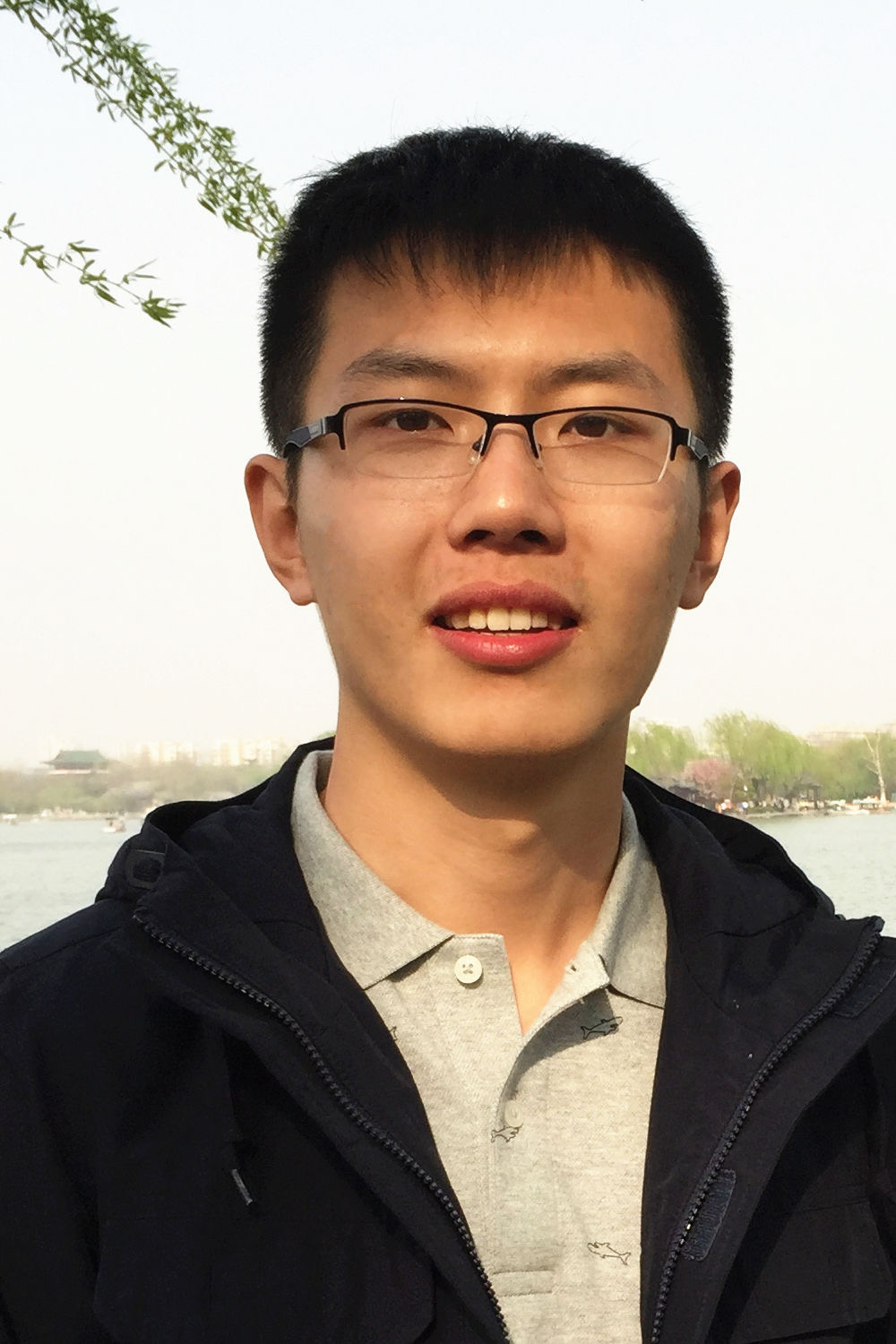}}]
{Yuning Jiang} received the B.Sc.
degree in electronic engineering from Shandong University, Jinan, China, in 2014, and the Ph.D. degree in information engineering from ShanghaiTech University, Shanghai, China, and the University of Chinese Academy of Sciences, Beijing, China, in 2020. 
He has ever been a Visiting Scholar with the
University of California at Berkeley (UC Berkeley), Berkeley, CA, USA, the University of Freiburg, Freiburg im Breisgau, Germany, and Technische Universit\"at Ilmenau (TU Ilmenau), Ilmenau, Germany, during his Ph.D. study. He is currently a Post-Doctoral Researcher with the Automatic Control Laboratory at the EPFL in Switzerland. His research focuses on learning- and optimization-based policy for operating complex systems such as nonlinear autonomous systems (e.g., autonomous vehicles, robotics and smart buildings), and large-scale multi-agent systems (e.g., power and energy systems, IoT and traffic networks).
\end{IEEEbiography}

\begin{IEEEbiography}[{\includegraphics[width=1in,height=1.25in,clip,keepaspectratio]{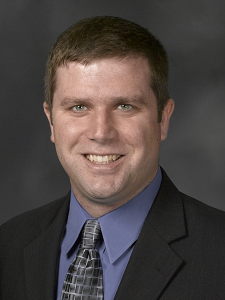}}]
{Daniel F. Opila} (SM) received the B.S. and M.S. degrees from the Massachusetts Institute of Technology, Cambridge, MA, USA, in 2002 and 2003, respectively, and the Ph.D. degree from the University of Michigan, Ann Arbor, MI, USA, in 2010.  He is currently an Associate Professor of Electrical and Computer Engineering at the United States Naval Academy, Annapolis, MD, USA. He has previously worked in various engineering positions at GE Power Conversion, Ford Motor Company, Orbital Sciences Corporation, and Bose Corporation. He specializes in optimal control of energy systems, including hybrid vehicles, naval power systems, power converters, and renewables.
\end{IEEEbiography}

\begin{IEEEbiography}[{\includegraphics[width=1in,height=1.25in,clip,keepaspectratio]{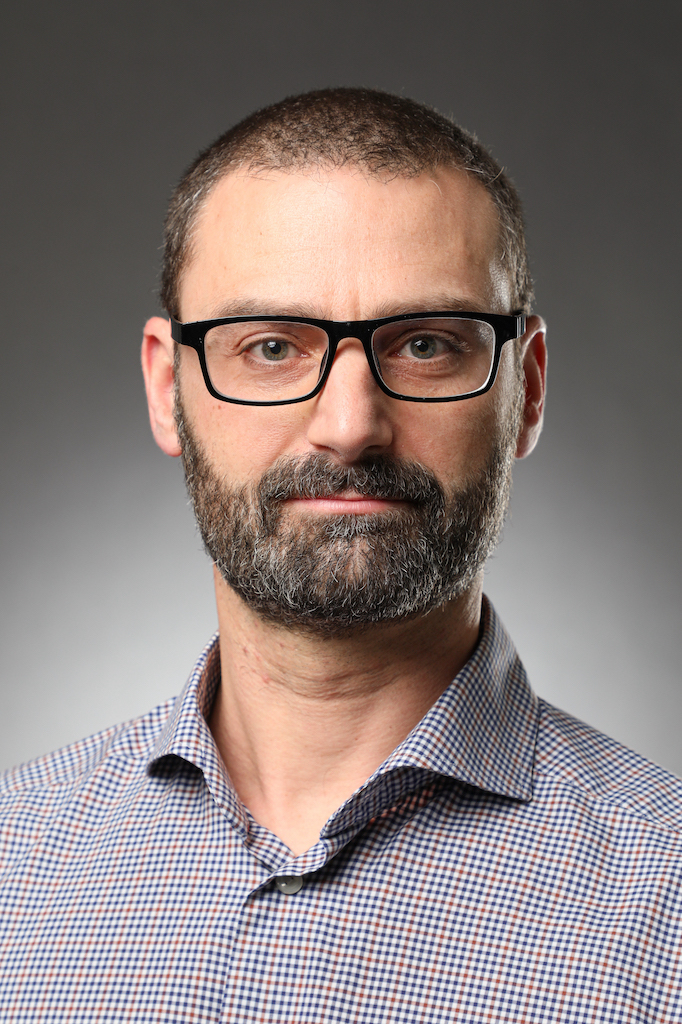}}]
{Colin N. Jones} has been an Associate Professor in the Automatic Control Laboratory at the EPFL in Switzerland since 2017 and an assistant professor from 2011. He was a Senior Researcher at the Automatic Control Lab at ETH Zurich until 2010 and obtained a Ph.D. in 2005 from the University of Cambridge for his work on polyhedral computational methods for constrained control. Prior to that, he was at the University of British Columbia in Canada, where he took his bachelor and master degrees in Electrical Engineering and Mathematics. He is the author or coauthor of more than 200 publications and was awarded an ERC starting grant to study the optimal control of building networks.
His current research interests are in the areas of high-speed predictive control and optimization, as well as the control of green energy generation, distribution and management.
\end{IEEEbiography}
\end{document}

%% file: grp_fig.tex
    
    
    
\tikzset{
    vertex/.style={
        fill,
        shape=circle,
        node distance=80pt},
    edge/.style={
        fill,
        opacity=.2,
        fill opacity=.5,
        line cap=round,
        line join=round,
        line width=20pt},
    elabel/.style={
        fill,
        shape=circle,
        node distance=30pt,
        opacity = .2}
}

\resizebox{\linewidth}{!}{
\begin{tikzpicture}


\node[vertex,label={[font = \Large]above:{$u_0$}}] (v2) {};

\node[vertex,right of=v2,label={[font = \Large]above:{$u_1$}}] (v3) {};

\node[vertex,right of=v3,label={[font = \Large]above:{$\dots$}}] (v4) {};

\node[vertex,right of=v4,label={[font = \Large]above:{$u_{N-1}$}}] (v5) {};


\node[vertex,below of=v2,label={[font = \Large]above:{$x_0$}}] (v6) {};

\node[vertex,right of=v6,label={[font = \Large]above:{$x_1$}}] (v7) {};

\node[vertex,right of=v7,label={[font = \Large]above:{$x_2$}}] (v8) {};

\node[vertex,right of=v8,label={[font = \Large]above:{$\dots$}}] (v9) {};

\node[vertex,right of=v9,label={[font = \Large]above:{$x_N$}}] (v10) {};


\draw[edge,color=green] (v2) -- (v7);

\draw[edge,color=blue] (v3) -- (v8);

\draw[edge,color=violet] (v5) -- (v10);

\end{tikzpicture}
}

%% file: building_plot.tex
\definecolor{MyRed}{HTML}{e6550d}
\definecolor{MyBlue}{rgb}{0.20, 0.6, 0.78}
\definecolor{MyGreen}{rgb}{0.4,0.8,0.4}
\begin{figure}[htbp!]
    \centering
    \begin{tikzpicture}
    \begin{axis}[xmin=0, xmax=22,
    ymin=10, ymax= 12,
    enlargelimits=false,
    clip=true,
    grid=major,
    mark size=0.5pt,
    width=1\linewidth,
    height=0.45\linewidth,ylabel = {Outdoor Temperature ($^\circ C$)},
    legend style={
    	font=\footnotesize,
    	draw=none,
		at={(0.5,1.03)},
        anchor=south
    },
    legend columns=5,
    label style={font=\scriptsize},
    ylabel style={at={(axis description cs:-0.07,0.5)}},
    xlabel style={at={(axis description cs:0.5,-0.05)}},
    ticklabel style = {font=\scriptsize}]
    
    \pgfplotstableread{data/building_vio.dat}{\dat};
    \addplot+ [ultra thick, mark=none, mark options={fill=white, scale=1.2},MyRed] table [x={t}, y={TOutdoor}] {\dat};
    \addlegendentry{recorded weather};
    \addplot+ [thick, mark=none, mark options={fill=white, scale=1.2},MyBlue] table [x={t}, y={TOutdoorPred}] {\dat};
    \addlegendentry{weather forecast}
    \end{axis}
    \end{tikzpicture}\\
    (a)\\
    \begin{tikzpicture}
    \begin{axis}[xmin=0, xmax=22,
    ymin=0, ymax= 0.3,
    enlargelimits=false,
    clip=true,
    grid=major,
    mark size=0.5pt,
    width=1\linewidth,
    height=0.45\linewidth,ylabel = {Solar radiation ($kW/m^2$)},
    legend style={
    	font=\footnotesize,
    	draw=none,
		at={(0.5,1.03)},
        anchor=south
    },
    legend columns=5,
    label style={font=\scriptsize},
    ylabel style={at={(axis description cs:-0.07,0.5)}},
    xlabel style={at={(axis description cs:0.5,-0.05)}},
    ticklabel style = {font=\scriptsize}]
    
    \pgfplotstableread{data/building_vio.dat}{\dat};
    \addplot+ [ultra thick, mark=none, mark options={fill=white, scale=1.2},MyRed] table [x={t}, y expr=\thisrow{SolOutdoor}/1000] {\dat};
    \addlegendentry{recorded weather};
    \addplot+ [thick, mark=none, mark options={fill=white, scale=1.2},MyBlue] table [x={t}, y expr=\thisrow{SolOutdoorPred}/1000] {\dat};
    \addlegendentry{weather forecast}
    \end{axis}
    \end{tikzpicture}\\
(b)\\
    \begin{tikzpicture}
    \begin{axis}[xmin=0, xmax=22,
    ymin=019.9, ymax= 24.5,
    enlargelimits=false,
    clip=true,
    grid=major,
    mark size=0.5pt,
    width=1.06\linewidth,
    height=0.6\linewidth,ylabel = {Simulated Indoor Temp. ($^\circ C$)},xlabel= {time ($h$)},
    legend style={
    	font=\footnotesize,
    	draw=none,
		at={(0.5,1.03)},
        anchor=south
    },
    legend columns=3,
    label style={font=\scriptsize},
    ylabel style={at={(axis description cs:-0.04,0.5)}},
    xlabel style={at={(axis description cs:0.5,-0.08)}},
    ticklabel style = {font=\scriptsize}]
    
    \pgfplotstableread{data/building_vio.dat}{\dat};
    \addplot+ [thick, mark=none, mark options={fill=white, scale=1.2},MyGreen] table [x={t}, y={y1}] {\dat};
    \addlegendentry{Alg~\ref{alg::solver}: 1};
     \addplot+ [thick, mark=none, mark options={fill=white, scale=1.2},MyBlue] table [x={t}, y={y2}] {\dat};
    \addlegendentry{Alg~\ref{alg::solver}: 2};
     \addplot+ [thick, mark=none, mark options={fill=white, scale=1.2},MyRed] table [x={t}, y={y3}] {\dat};
    \addlegendentry{Alg~\ref{alg::solver}: 3};
     \addplot+ [thick, mark=none, mark options={fill=white, scale=1.2},purple] table [x={t}, y={y4}] {\dat};
    \addlegendentry{Alg~\ref{alg::solver}: 4};
    \addplot+ [ultra thick, dashed, mark=none, mark options={fill=white, scale=1.2},MyGreen] table [x={t}, y={yac1}] {\dat};
    \addlegendentry{\texttt{acados: 1}};
    \addplot+ [ultra thick, dashed, mark=none, mark options={fill=white, scale=1.2},MyBlue] table [x={t}, y={yac2}] {\dat};
    \addlegendentry{\texttt{acados: 2}};
    \addplot+ [ultra thick, dashed, mark=none, mark options={fill=white, scale=1.2},MyRed] table [x={t}, y={yac3}] {\dat};
    \addlegendentry{\texttt{acados: 3}};
    \addplot+ [ultra thick, dashed, mark=none, mark options={fill=white, scale=1.2},purple] table [x={t}, y={yac4}] {\dat};
    \addlegendentry{\texttt{acados: 4}};
    \addplot+ [thick, dashed, mark=none, mark options={fill=white, scale=1.2},black,forget plot] table [x={t}, y={ymax}] {\dat};
    \addplot+ [thick, dashed, mark=none, mark options={fill=white, scale=1.2},black] table [x={t}, y={ymin}] {\dat};
    \addlegendentry{constraints}
    \end{axis}
    \end{tikzpicture}\\
    (c)
    \caption{\change{\label{fig:building}Case study of sudden indoor temperature drop: (a) - (b): a sample of the recorded and forecast weather condition, (a) outdoor temperature, (b) solar radiation. The forecast is used as the nominal weather in the NMPC problem, while recorded weather is used for the simulation of building dynamics. (c): simulation of indoor temperature of different rooms (room index depicted in Figure~\ref{fig:multizone_scheme}).}}
\end{figure}

%% file: current_curve.tex
\definecolor{MyRed}{HTML}{e6550d}
\definecolor{MyBlue}{rgb}{0.20, 0.6, 0.78}
\definecolor{MyGreen}{rgb}{0.4,0.8,0.4}
\begin{figure}[htbp!]
    \centering
    \begin{tikzpicture}
    \begin{axis}[xmin=0, xmax=200,
    ymin=0, ymax= 4.3,
    enlargelimits=false,
    clip=true,
    grid=major,
    mark size=0.5pt,
    ytick distance = 0.5,
    width=1\linewidth,
    height=.8\linewidth,xlabel={Speed $[\textrm{rad/s}]$},ylabel = {Torque $[\mathrm{Nm}]$},
    legend style={
    	font=\footnotesize,
    	draw=none,
		at={(0.5,1.03)},
        anchor=south
    },
    legend columns=5,
    label style={font=\scriptsize},
    ylabel style={at={(axis description cs:-0.05,0.5)}},
    xlabel style={at={(axis description cs:0.5,-0.05)}},
    ticklabel style = {font=\scriptsize}]
    
    \pgfplotstableread{data/torque_curve.dat}{\dat};
    \addplot+ [thick, solid,mark=none, mark options={fill=white, scale=1.2},YellowOrange] table [x={x}, y ={y_1}] {\dat};
    \addplot+ [thick, solid,mark=none, mark options={fill=white, scale=1.2},Apricot] table [x={x}, y ={y_2}] {\dat};
    \addplot+ [thick, solid,mark=none, mark options={fill=white, scale=1.2},Brown] table [x={x}, y ={y_3}] {\dat};
    \addplot+ [thick, solid,mark=none, mark options={fill=white, scale=1.2},DarkOrchid] table [x={x}, y ={y_4}] {\dat};
    \addplot+ [thick, solid,mark=none, mark options={fill=white, scale=1.2},JungleGreen] table [x={x}, y ={y_5}] {\dat};
    \addplot+ [thick,solid, mark=none, mark options={fill=white, scale=1.2},Mulberry] table [x={x}, y ={y_6}] {\dat};
    \addplot+ [thick,solid, mark=none, mark options={fill=white, scale=1.2},PineGreen] table [x={x}, y ={y_7}] {\dat};
    \addplot+ [thick,solid, mark=none, mark options={fill=white, scale=1.2}SpringGreen] table [x={x}, y ={y_8}] {\dat};
    \addplot+ [thick,solid, mark=none, mark options={fill=white, scale=1.2},WildStrawberry] table [x={x}, y ={y_9}] {\dat};
    \addplot+ [thick,solid, mark=none, mark options={fill=white, scale=1.2},BlueGreen] table [x={x}, y ={y_10}] {\dat};
    \addplot+ [thick,solid, mark=none, mark options={fill=white, scale=1.2},CornflowerBlue] table [x={x}, y ={y_11}] {\dat};
    \addplot+ [thick,solid, mark=none, mark options={fill=white, scale=1.2},Gray] table [x={x}, y ={y_12}] {\dat};
    \addplot+ [thick,solid, mark=none, mark options={fill=white, scale=1.2},NavyBlue] table [x={x}, y ={y_13}] {\dat};
    \addplot+ [thick,solid, mark=none, mark options={fill=white, scale=1.2},Violet] table [x={x}, y ={y_14}] {\dat};
    \addplot+ [thick,solid, mark=none, mark options={fill=white, scale=1.2},Periwinkle] table [x={x}, y ={y_15}] {\dat};
    
    \node[] at (axis cs: 4.8,0.1) {\scriptsize 0.2};
    \node[] at (axis cs: 4.8,0.45) {\scriptsize 0.4};
    \node[] at (axis cs: 4.8,0.7) {\scriptsize 0.6};
    \node[] at (axis cs: 4.8,1.0) {\scriptsize 0.8};
    \node[] at (axis cs: 4.8,1.25) {\scriptsize 1.0};
    \node[] at (axis cs: 4.8,1.5) {\scriptsize 1.2};
    \node[] at (axis cs: 4.8,1.8) {\scriptsize 1.4};
    \node[] at (axis cs: 4.8,2.05) {\scriptsize 1.6};
    \node[] at (axis cs: 4.8,2.3) {\scriptsize 1.8};
    \node[] at (axis cs: 4.8,2.55) {\scriptsize 2.0};
    \node[] at (axis cs: 4.8,2.8) {\scriptsize 2.2};
    \node[] at (axis cs: 4.8,3.05) {\scriptsize 2.4};
    \node[] at (axis cs: 4.8,3.3) {\scriptsize 2.6};
    \node[] at (axis cs: 4.8,3.55) {\scriptsize 2.8};
    \node[] at (axis cs: 4.8,3.9) {\scriptsize 3.0};
    
    \end{axis}
    \end{tikzpicture}\\
    (a)
    \begin{tikzpicture}
    \begin{axis}[xmin=0, xmax=200,
    ymin=0, ymax= 6,
    enlargelimits=false,
    clip=true,
    grid=major,
    mark size=0.5pt,
    ytick distance = 0.5,
    width=1\linewidth,
    height=.8\linewidth,xlabel={Speed $[\textrm{rad/s}]$},ylabel = {Armature current $[\mathrm{A}]$},
    legend style={
    	font=\footnotesize,
    	draw=none,
		at={(0.5,1.03)},
        anchor=south
    },
    legend columns=5,
    label style={font=\scriptsize},
    ylabel style={at={(axis description cs:-0.05,0.5)}},
    xlabel style={at={(axis description cs:0.5,-0.05)}},
    ticklabel style = {font=\scriptsize}]
    
    \pgfplotstableread{data/current_curve.dat}{\dat};
    \addplot+ [thick, solid,mark=none, mark options={fill=white, scale=1.2},YellowOrange] table [x={x}, y ={y_1}] {\dat};
    \addplot+ [thick, solid,mark=none, mark options={fill=white, scale=1.2},Apricot] table [x={x}, y ={y_2}] {\dat};
    \addplot+ [thick, solid,mark=none, mark options={fill=white, scale=1.2},Brown] table [x={x}, y ={y_3}] {\dat};
    \addplot+ [thick, solid,mark=none, mark options={fill=white, scale=1.2},DarkOrchid] table [x={x}, y ={y_4}] {\dat};
    \addplot+ [thick, solid,mark=none, mark options={fill=white, scale=1.2},JungleGreen] table [x={x}, y ={y_5}] {\dat};
    \addplot+ [thick,solid, mark=none, mark options={fill=white, scale=1.2},Mulberry] table [x={x}, y ={y_6}] {\dat};
    \addplot+ [thick,solid, mark=none, mark options={fill=white, scale=1.2},PineGreen] table [x={x}, y ={y_7}] {\dat};
    \addplot+ [thick,solid, mark=none, mark options={fill=white, scale=1.2}SpringGreen] table [x={x}, y ={y_8}] {\dat};
    \addplot+ [thick,solid, mark=none, mark options={fill=white, scale=1.2},WildStrawberry] table [x={x}, y ={y_9}] {\dat};
    \addplot+ [thick,solid, mark=none, mark options={fill=white, scale=1.2},BlueGreen] table [x={x}, y ={y_10}] {\dat};
    \addplot+ [thick,solid, mark=none, mark options={fill=white, scale=1.2},CornflowerBlue] table [x={x}, y ={y_11}] {\dat};
    \addplot+ [thick,solid, mark=none, mark options={fill=white, scale=1.2},Gray] table [x={x}, y ={y_12}] {\dat};
    \addplot+ [thick,solid, mark=none, mark options={fill=white, scale=1.2},NavyBlue] table [x={x}, y ={y_13}] {\dat};
    \addplot+ [thick,solid, mark=none, mark options={fill=white, scale=1.2},Violet] table [x={x}, y ={y_14}] {\dat};
    \addplot+ [thick,solid, mark=none, mark options={fill=white, scale=1.2},Periwinkle] table [x={x}, y ={y_15}] {\dat};
    \addplot+ [ultra thick,dashed, mark=none, mark options={fill=white, scale=1.2},MyRed] table [x={x}, y ={ymax}] {\dat};
    
    \node[] at (axis cs: 190,4.8) {\scriptsize 0.2};
    \node[] at (axis cs: 190,4) {\scriptsize 0.4};
    \node[] at (axis cs: 190,3.2) {\scriptsize 0.6};
    \node[] at (axis cs: 190,2.2) {\scriptsize 0.8};
    \node[] at (axis cs: 190,1.3) {\scriptsize 1.0};
    \node[] at (axis cs: 190,.5) {\scriptsize 1.2};
    \node[] at (axis cs: 175,.2) {\scriptsize 1.4};
    \node[] at (axis cs: 150,.2) {\scriptsize 1.6};
    \node[] at (axis cs: 135,.2) {\scriptsize 1.8};
    \node[] at (axis cs: 125,.2) {\scriptsize 2.0};
    \node[] at (axis cs: 112,.2) {\scriptsize 2.2};
    \node[] at (axis cs: 100,.2) {\scriptsize ...};
    \node[] at (axis cs: 80.8,.2) {\scriptsize 3.0};
    
    \end{axis}
    \end{tikzpicture}\\
    (b)
    \caption{\newchange{Torque and current} curves of the DC motor. Top: Torque-Speed curve at different field currents (indicated in Amps on the left end of each straight line). Bottom: Armature current as a function of speed for different field currents (indicated in Amps on the right end of each straight line). To avoid armature overheating, the armature current should stay below the 3 A thick red dashed line in long term operation.}
    \label{fig:current_curve}
\end{figure}

%% file: motor_plot_sim.tex
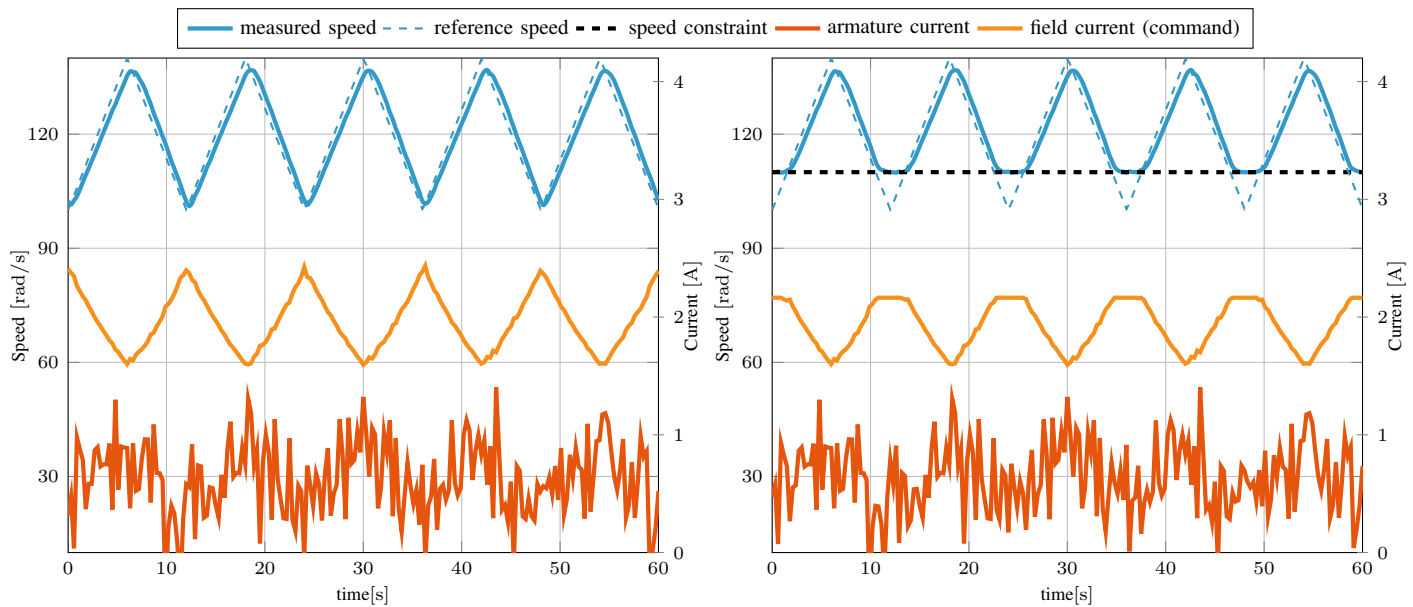
\begin{figure*}[t]
    \centering
    \begin{tikzpicture}
    \begin{groupplot}[
        legend columns=5,
        legend style={
    	font=\footnotesize},
        group style=
            {columns = 2,horizontal sep = 4.3em,
            group name=plots}]
    \nextgroupplot[xmin=0, xmax=60,
    ymin=10, ymax= 140,
    ytick distance = 30,
    enlargelimits=false,
    clip=true,
    grid=major,
    mark size=0.5pt,
    width=.52\linewidth,
    height=0.45\linewidth,ylabel = {Speed $[\mathrm{rad/s}]$} ,xlabel= {time$[\mathrm{s}]$},
    legend to name=grouplegend,
    label style={font=\scriptsize},
    ticklabel style = {font=\scriptsize},
    ylabel style={at={(axis description cs:-0.05,0.5)}},
    xlabel style={at={(axis description cs:0.5,-0.05)}}]
    \pgfplotstableread{data/motor_sim.dat}{\dat}
    \addplot+ [ultra thick, mark=none, mark options={fill=white, scale=1.2},MyBlue,smooth] table [x={t}, y={w}] {\dat};
    \addlegendentry{measured speed}
    \addplot+ [thick, dashed, mark=none, mark options={fill=white, scale=1.2},MyBlue] table [x={t}, y={ref}] {\dat};
    \addlegendentry{reference speed}
    
     \addlegendimage{line legend,ultra thick, dashed, mark=none, mark options={fill=white, scale=1.2},black}
    \addlegendentry{speed constraint}  
    
    \addlegendimage{line legend,ultra thick,  mark=none, mark options={fill=white, scale=1.2},MyRed};
    \addlegendentry{armature current};
    \addlegendimage{line legend,ultra thick,  mark=none, mark options={fill=white, scale=1.2},BurntOrange};
    \addlegendentry{field current (command)};
    
    \nextgroupplot[xmin=0, xmax=60,
    ymin=10, ymax= 140,
    enlargelimits=false,
    ytick distance = 30,
    clip=true,
    grid=major,
    mark size=0.5pt,
    width=.52\linewidth,
    height=0.45\linewidth,ylabel = {Speed $[\mathrm{rad/s}]$} ,xlabel= {time$[\mathrm{s}]$},
    label style={font=\scriptsize},
    ticklabel style = {font=\scriptsize},
    ylabel style={at={(axis description cs:-0.05,0.5)}},
    xlabel style={at={(axis description cs:0.5,-0.05)}}]
    \pgfplotstableread{data/motor_sim_clipped.dat}{\dat}
    \addplot+ [ultra thick, mark=none, mark options={fill=white, scale=1.2},MyBlue,smooth] table [x={t}, y={w}] {\dat};
    \addplot+ [ultra thick, dashed, mark=none, mark options={fill=white, scale=1.2},black] table [x={t}, y={wmin}] {\dat};
    \addplot+ [thick, dashed, mark=none, mark options={fill=white, scale=1.2},MyBlue] table [x={t}, y={ref}] {\dat};

    \end{groupplot}

    \begin{groupplot}
    [group style=
            {columns = 2,horizontal sep = 4.3em},
    xtick=\empty, axis line style=transparent,
        ytick align=outside,
        ytick pos=right,
        axis y line=right,]
    
    \nextgroupplot[xmin=0, xmax=60,
    ymin=0, ymax= 4.2,
    enlargelimits=false,
    clip=true,
    mark size=0.5pt,
    width=.52\linewidth,
    height=0.45\linewidth,ylabel = {Current $[\mathrm{A}]$} ,
    label style={font=\scriptsize},
    ticklabel style = {font=\scriptsize},
    ylabel style={at={(axis description cs:0.91,0.5)}},
    xlabel style={at={(axis description cs:0.5,0.05)}}]
    \pgfplotstableread{data/motor_sim.dat}{\dat}
    \addplot+ [ultra thick,  mark=none, mark options={fill=white, scale=1.2},MyRed] table [x={t}, y={Iarm}] {\dat};
    \addplot+ [ultra thick,  mark=none, mark options={fill=white, scale=1.2},BurntOrange] table [x={t}, y={Ifield}] {\dat};

    \nextgroupplot[xmin=0, xmax=60,
    ymin=0, ymax= 4.2,
    enlargelimits=false,
    clip=true,
    mark size=0.5pt,
    width=.52\linewidth,
    height=0.45\linewidth,ylabel = {Current $[\mathrm{A}]$} ,
    label style={font=\scriptsize},
    ticklabel style = {font=\scriptsize},
    ylabel style={at={(axis description cs:.91,0.5)}},
    xlabel style={at={(axis description cs:0.5,-0.05)}}]
    \pgfplotstableread{data/motor_sim_clipped.dat}{\dat}
    \addplot+ [ultra thick,  mark=none, mark options={fill=white, scale=1.2},MyRed] table [x={t}, y={Iarm}] {\dat};
    \addplot+ [ultra thick,  mark=none, mark options={fill=white, scale=1.2},BurntOrange] table [x={t}, y={Ifield}] {\dat};

    \end{groupplot}
    \path (group c1r1.north east) -- node[above]{\ref*{grouplegend}} (group c2r1.north west);
    
    \end{tikzpicture}
   
    \caption{\change{Hardware in the loop simulation} of the motor control with the C2000 microcontroller. Left: speed constraint not active. Right: speed constraint \change{$[110,180]$ rad/s} active \change{(black dashed line)}. Raw signals are plotted.}
    \label{fig:motor_sim_result}
\end{figure*}

%% file: motor_plot.tex
\begin{figure*}[t]
    \centering
    \begin{tikzpicture}
    \begin{groupplot}[
        legend columns=5,
        legend style={
    	font=\footnotesize},
        group style=
            {columns = 2,horizontal sep = 4.3em,
            group name=plots}]
    \nextgroupplot[xmin=0, xmax=60,
    ymin=10, ymax= 140,
    ytick distance = 30,
    enlargelimits=false,
    clip=true,
    grid=major,
    mark size=0.5pt,
    width=.52\linewidth,
    height=0.45\linewidth,ylabel = {Speed $[\mathrm{rad/s}]$} ,xlabel= {time$[\mathrm{s}]$},
    legend to name=grouplegend,
    label style={font=\scriptsize},
    ticklabel style = {font=\scriptsize},
    ylabel style={at={(axis description cs:-0.05,0.5)}},
    xlabel style={at={(axis description cs:0.5,-0.05)}}]
    \pgfplotstableread{data/motor_test1.dat}{\dat}
    \addplot+ [ultra thick, mark=none, mark options={fill=white, scale=1.2},MyBlue,smooth] table [x={t}, y={w}] {\dat};
    \addlegendentry{measured speed}
    \pgfplotstableread{data/motor_ref_test1.dat}{\dat}
    \addplot+ [thick, dashed, mark=none, mark options={fill=white, scale=1.2},MyBlue] table [x={t}, y={ref}] {\dat};
    \addlegendentry{reference speed}
    
     \addlegendimage{line legend,ultra thick, dashed, mark=none, mark options={fill=white, scale=1.2},black}
    \addlegendentry{speed constraint}  
    
    \addlegendimage{line legend,ultra thick,  mark=none, mark options={fill=white, scale=1.2},MyRed};
    \addlegendentry{armature current};
    \addlegendimage{line legend,ultra thick,  mark=none, mark options={fill=white, scale=1.2},BurntOrange};
    \addlegendentry{field current (command)};
    
    \nextgroupplot[xmin=0, xmax=60,
    ymin=10, ymax= 140,
    enlargelimits=false,
    ytick distance = 30,
    clip=true,
    grid=major,
    mark size=0.5pt,
    width=.52\linewidth,
    height=0.45\linewidth,ylabel = {Speed $[\mathrm{rad/s}]$} ,xlabel= {time$[\mathrm{s}]$},
    label style={font=\scriptsize},
    ticklabel style = {font=\scriptsize},
    ylabel style={at={(axis description cs:-0.05,0.5)}},
    xlabel style={at={(axis description cs:0.5,-0.05)}}]
    \pgfplotstableread{data/motor_test2.dat}{\dat}
    \addplot+ [ultra thick, mark=none, mark options={fill=white, scale=1.2},MyBlue,smooth] table [x={t}, y={w}] {\dat};
    \addplot+ [ultra thick, dashed, mark=none, mark options={fill=white, scale=1.2},black] table [x={t}, y={wmin}] {\dat};
    \pgfplotstableread{data/motor_ref_test2.dat}{\dat}
    \addplot+ [thick, dashed, mark=none, mark options={fill=white, scale=1.2},MyBlue] table [x={t}, y={ref}] {\dat};

    \end{groupplot}

    \begin{groupplot}
    [group style=
            {columns = 2,horizontal sep = 4.3em},
    xtick=\empty, axis line style=transparent,
        ytick align=outside,
        ytick pos=right,
        axis y line=right,]
    
    \nextgroupplot[xmin=0, xmax=60,
    ymin=0, ymax= 4.2,
    enlargelimits=false,
    clip=true,
    mark size=0.5pt,
    width=.52\linewidth,
    height=0.45\linewidth,ylabel = {Current $[\mathrm{A}]$} ,
    label style={font=\scriptsize},
    ticklabel style = {font=\scriptsize},
    ylabel style={at={(axis description cs:0.91,0.5)}},
    xlabel style={at={(axis description cs:0.5,0.05)}}]
    \pgfplotstableread{data/motor_test1.dat}{\dat}
    \addplot+ [ultra thick,  mark=none, mark options={fill=white, scale=1.2},MyRed] table [x={t}, y={Iarm}] {\dat};
    \addplot+ [ultra thick, mark=none, mark options={fill=white, scale=1.2},BurntOrange,draw opacity = 0.8] table [x={t}, y={Icmd}] {\dat};

    \nextgroupplot[xmin=0, xmax=60,
    ymin=0, ymax= 4.2,
    enlargelimits=false,
    clip=true,
    mark size=0.5pt,
    width=.52\linewidth,
    height=0.45\linewidth,ylabel = {Current $[\mathrm{A}]$} ,
    label style={font=\scriptsize},
    ticklabel style = {font=\scriptsize},
    ylabel style={at={(axis description cs:.91,0.5)}},
    xlabel style={at={(axis description cs:0.5,-0.05)}}]
    \pgfplotstableread{data/motor_test2.dat}{\dat}
    \addplot+ [ultra thick,  mark=none, mark options={fill=white, scale=1.2},MyRed] table [x={t}, y={Iarm}] {\dat};
    \addplot+ [ultra thick, mark=none, mark options={fill=white, scale=1.2},BurntOrange,draw opacity = 0.8] table [x={t}, y={Icmd}] {\dat};

    \end{groupplot}
    \path (group c1r1.north east) -- node[above]{\ref*{grouplegend}} (group c2r1.north west);
    
    \end{tikzpicture}
   
    \caption{Hardware testing of the proposed MPC algorithm with a real motor controlled by a C2000 microcontroller. Left: speed constraint not active. Right: speed constraint $[110,180]$ rad/s active \change{(black dashed line)}. The measured signals are post-processed with a low-pass zero-phase forward-reverse filter.}
    \label{fig:motor_hardwareMPC_result}
\end{figure*}
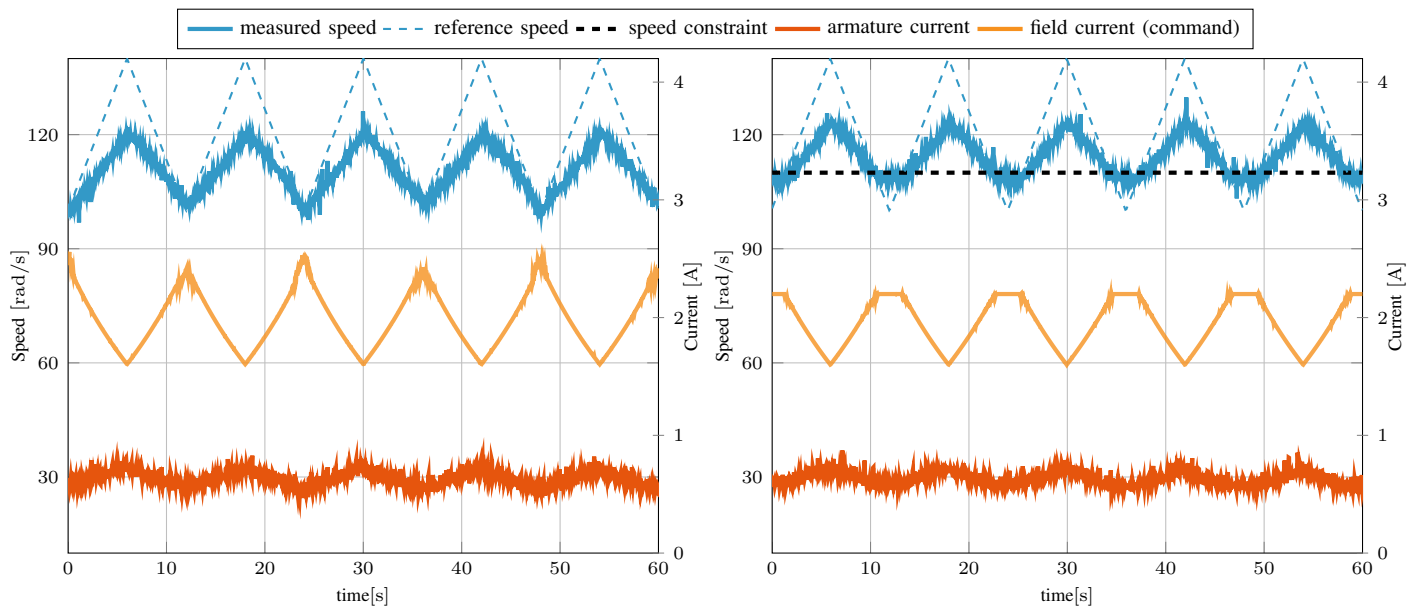